\documentclass[a4paper]{article}
\usepackage[a4paper]{geometry}
\usepackage{amsfonts}
\usepackage{latexsym}
\usepackage{amsmath}
\usepackage{amssymb}
\usepackage{mathbbol}
\usepackage{graphicx}
\author{Friedrich Hubalek\thanks{Vienna University of Technology, Austria}
\and Petra Posedel\thanks{Department of Mathematics, Faculty of
Economics \& Business Zagreb, Croatia\newline The second author
gratefully acknowledges financial support from the Austrian
Science Fund (FWF) under grants P18022 and START prize Y328.}}
\title{Joint analysis and estimation of stock prices and trading volume in
Barndorff-Nielsen and Shephard stochastic volatility models}
\def\proof{{\textbf Proof:~}}
\def\qed{\mbox{}\hfill$\Box$}
\date{}
\def\Levy{L\'evy}

\def\asto{\stackrel{a.s.}{\longrightarrow}}
\def\inlawto{\stackrel{\mathcal D}{\longrightarrow}}

\def\Var{\mbox{Var}}
\newtheorem{theorem}{Theorem}
\newtheorem{remark}{Remark}
\newtheorem{lemma}{Lemma}

\newtheorem{proposition}{Proposition}

\newtheorem{assumption}{Assumption}

\newcommand{\dis}{\displaystyle}
\newcommand{\lijepof}{\mathcal{F}}

\begin{document}
\maketitle
\begin{abstract}
We introduce a variant of the Barndorff-Nielsen and Shephard
stochastic volatility model where the non Gaussian
Ornstein-Uhlenbeck process describes some measure of trading
intensity like trading volume or number of trades instead of
unobservable instantaneous variance. We develop an explicit
estimator based on martingale estimating functions in a bivariate
model that is not a diffusion, but admits jumps. It is assumed
that both the quantities are observed on a discrete grid of fixed
width, and the observation horizon tends to infinity. We show that
the estimator is consistent and asymptotically normal and give
explicit expressions of the asymptotic covariance matrix. Our
method is illustrated by a finite sample experiment and a
statistical analysis on the International Business Machines
Corporation (IBM) stock from the New York Stock Exchange (NYSE)
and the Microsoft Corporation (MSFT) stock from Nasdaq during a
history of five years.
\end{abstract}
\subsubsection*{KEYWORDS:} Martingale estimating functions,
stochastic volatility models with jumps, consistency and
asymptotic normality, trading intensity
\section{Introduction}
In \cite{BNS} Barndorff-Nielsen and Shephard introduced a class of
stochastic volatility models in continuous time, where the
instantaneous variance follows an Ornstein-Uhlenbeck type process
driven by an increasing \Levy{} process. BNS-models, as we will
call them from now on, allow flexible modelling, capture many
stylized facts of financial time series, and yet are of great
analytical tractability. Those models have been studied from
various points of view in mathematical finance and related fields.
Unfortunately, it seems that statistical estimation of the model
is the most difficult problem, and most of the work in that area
is focused on computationally intensive methods. In \cite{HP2007}
an explicit estimator based on martingale estimating functions was
developed under the assumptions that returns and volatility are
observed. That paper contains also further references on BNS
models, martingale estimating functions, estimating discretely
observed diffusions models, etc. The literature on estimation for
discretely observed diffusions is vast, a few references are
\cite{di2,di6,di7,di10,di11,di13}. In particular, the martingale
estimating function approach is used, developed and studied for
example in \cite{SO99}. In the diffusion setting the major
difficulty is that the transition probabilities are not known and
are difficult to compute. In contrast to that, the characteristic
function of the transition probability is known in closed form for
many BNS models and the transition probability can be computed
with Fourier methods with high precision.

In practice volatility is not observed, but many researchers,
including, for example, \cite{JMK87,GRT92,JKL94} have established
a connection between volatility and different measures of trading
intensity, such as traded volume or number of trades. In
particular, \cite{Lin2007} gives a first application of this
approach to BNS models. We take up this idea and combine it with
the martingale estimating function approach. Measures of trading
intensity contain much information about the volatility. We
identify the volatility with a multiple of some measure of trading
intensity, in this paper the daily traded volume. In doing so, our
bivariate time series is given by the logarithmic returns and
trading volume which are both observable quantities. We explore
the joint distribution of logarithmic returns~$X$ and the
instantaneous trading volume/number of trades~$\tau$. The joint
conditional moment-generating function of $(X,\tau)$ is known in
closed form and thus we obtain closed form expressions for the
joint conditional moments up to any desired order. This yields a
sequence of martingale differences and  the martingale estimating
function approach is used. We employ then the large sample
properties, in particular the strong law of large numbers for
martingales and the martingale central limit theorem. In this way
we do not need ergodicity, mixing conditions, etc.

The contributions of the present paper are as follows: first we
develop a simple and explicit estimator for BNS models using a
martingale estimating function approach and identifying the
volatility with a multiple of trading volume. Secondly, we give
proofs of its consistency and asymptotic normality. In doing so we
compute explicitly the asymptotic covariance matrix. Thirdly, we
include numerical illustrations and apply our method on real data.

Since in this analysis we assume that the discrete time variance
process $V_i$ is proportional to the trading volume/number of
trades $\tau_i$, we are able to directly model the stochastic
volatility in asset price dynamics. Due to the analytical
tractability of BNS models, we can work with the exact dynamics
for discrete observations of the continuous time model. We want to
stress that our approach leads to simple and explicit formulas for
the estimator and its asymptotic covariance matrix, and no
simulation or other computer intensive methods are required.
Simulations are only used to illustrate the finite sample
performance in numerical experiments.
 Finally, we apply the method to real data and do a statistical
analysis on the International Business Machines Corporation (IBM)
stock from the New York Stock Exchange (NYSE) and the Microsoft
Corporation (MSFT) stock from Nasdaq during a volatile history of
five years.

The remainder of the paper is organized as follows: in
section~\ref{continuous} we describe the class of BNS models in
continuous time and present two concrete examples, the $\Gamma-$OU
and IG-OU model. In section~\ref{discrete} we introduce the
quantities observed in discrete time that are used for estimation.
 In section~\ref{main} we present
the estimating equations, their explicit solution which is our
estimator and its consistency and asymptotic normality are proven. 
In section~\ref{numerical} numerical illustrations are presented.
In section~\ref{empirical} we apply our results on daily data on
the IBM stock from NYSE and the MSFT stock from Nasdaq.
\section{The model}
\subsection{The continuous time model\label{continuous}}
\subsubsection{The general setting}
As in Barndorff-Nielsen and Shepard \cite{BNS}, we assume that the
price process of an asset $S$ is defined on some filtered
probability space $\left(\Omega,\mathcal F,(\mathcal
F_t)_{t\geq0},P\right)$ and is given by $S_t=S_0\exp(X_t)$ with
$S_0>0$ a constant. The process of logarithmic returns~$X$ and the
instantaneous trading volume/number of trades process~$\tau$
satisfy
\begin{equation}\label{returns}
dX(t)=(\mu+\beta
\tau(t-))dt+\sigma\sqrt{\tau(t-)}dW_\theta(t)+\rho
dZ_\lambda(t),\quad X(0)=0.
\end{equation}
and
\begin{equation}\label{dV}
d\tau(t)=-\lambda \tau(t-)dt+dZ_\lambda(t),\quad \tau(0)=\tau_0,
\end{equation}
where the parameters $\mu, \beta, \rho, \sigma$ and $\lambda$ are
real constants with $\lambda,\sigma>0.$ The process $W$ is a
standard Brownian motion, the process $Z$  is an increasing
L\'{e}vy process, and we define $Z_\lambda(t)=Z(\lambda t)$ for
notational simplicity. Adopting the terminology introduced by
Barndorff-Nielsen and Shepard, we will refer to $Z$ as the
\emph{background driving L\'{e}vy process} (BDLP). The Brownian
motion $W$ and the BDLP $Z$ are independent and $(\mathcal F_t)$
is assumed to be the usual augmentation of the filtration
generated by the pair $(W,Z_\lambda)$. The random variable
$\tau_0$ has a self-decomposable distribution corresponding to the
BDLP such that the process $\tau$ is strictly stationary and
\begin{equation}
E[\tau_0]=\zeta,\qquad \Var[\tau_0]=\eta.
\end{equation}
For our analysis we will assume that the instantaneous variance
process~$V$ is a constant time the trading volume/number of trades
$\tau.$ That is,
\begin{equation}\label{trades}dV(t)=\sigma^2\cdot d\tau(t),
\end{equation}
with $\sigma>0.$
\begin{remark}Equation (\ref{trades}) implies that the instantaneous variance of log returns
is a constant multiple of the trading volume/number of trades, and
trading volume/number of trades is modelled as an OU-type process.
\end{remark}
To shorten the notation we introduce the parameter vector
\begin{equation}
\theta=(\nu,\alpha,\lambda,\mu,\beta,\sigma,\rho)^\top,
\end{equation}
and the bivariate process
\begin{equation}
\mathbf X=(X,\tau).
\end{equation}
If the distribution of $\tau_0$ is from a particular class $D$
then $\mathbf{X}$ is called a BNS-DOU($\theta$) model.

The process $(X_t,\tau_t)_{t\geq0}$ is clearly Markovian.
\subsubsection{The $\Gamma$-OU model\label{Sec-GaOU}}
The $\Gamma$-OU model is obtained by constructing the BNS-model
with stationary gamma distribution,
$\tau_0\sim\Gamma(\nu,\alpha)$, where the parameters are $\nu>0$
and $\alpha>0$. The corresponding background driving \Levy{}
process~$Z$ is a compound Poisson processes with intensity $\nu$
and jumps from the exponential distribution with parameter
$\alpha$. Consequently both processes $Z$ and $\tau$ have a finite
number of jumps in any finite time interval.

For the $\Gamma$-OU model it is more convenient to work with the
parameters $\nu$ and $\alpha$. The connection to the generic
parameters used in our general development is given by
\begin{equation}\label{al-nu}
\zeta=\frac{\nu}{\alpha},
\qquad
\eta=\frac{\nu}{\alpha^2}.
\end{equation}
As the gamma distribution admits exponential moments we have
integer moments of all orders and our Assumption~\ref{moments}
below is satisfied.
\subsubsection{The IG-OU model}\label{Sec-IGOU}
The IG-OU model is obtained by constructing the BNS-model with
stationary inverse Gaussian distribution,
$\tau_0\sim(\delta,\gamma)$, with parameters $\delta>0$ and
$\gamma>0$.

The corresponding background driving \Levy{} process is the sum of
an IG($\delta/2,\gamma)$ process and an independent compound
Poisson process with intensity $\delta\gamma/2$ and jumps from an
$\Gamma(1/2,\gamma^2/2)$ distribution. Consequently both processes
$Z$ and $\tau$ have infinitely many jumps in any finite time
interval.

For the IG-OU model it is more convenient to work with the
parameters $\delta$ and $\gamma$. The connection to the generic
parameters used in our general development is given by
\begin{equation}\label{de-ga}
\zeta=\frac{\delta}{\gamma},
\qquad
\eta=\frac{\delta}{\gamma^3}.
\end{equation}
As the inverse Gaussian distribution admits exponential moments we have
integer moments of all orders and our Assumption~\ref{moments}
below is satisfied.
\subsection{Discrete observations\label{discrete}}
The following description is rather analogous to
\cite[Section~$2.2$]{HP2007}. The only (but important) exception
is the introduction of the parameter $\sigma$ in (\ref{returns2})
and (\ref{defxd}). We observe returns and the trading
volume/number of trades process on a discrete grid of points in
time
\begin{equation}
0=t_0<t_1<\ldots<t_n,
\end{equation}
which relates trading volume/number of trades and the
instantaneous variance of log returns. This implies
\begin{equation}\label{deft}
\tau(t_i)=\tau(t_{i-1})e^{-\lambda(t_i-t_{i-1})}
+\int_{t_{i-1}}^{t_i}e^{-\lambda(t_i-s)}dZ_\lambda(s).
\end{equation}
Using
\begin{equation}\label{defuv}
\tau_i:=\tau(t_i),\quad
U_i:=\int_{t_{i-1}}^{t_i}e^{-\lambda(t_i-s)}dZ_\lambda(s)
\end{equation}
we have that $(U_i)_{i\geq1}$ is a sequence of independent random
variables, and it is independent of~$\tau_0$. If the grid is
equidistant, then $(U_i)_{i\geq1}$ are iid. Observing the returns
$X$ on the grid we have
\begin{equation}\label{returns2}
\renewcommand{\arraystretch}{2}
\begin{array}{l}
X(t_i)-X(t_{i-1})=\mu(t_i-t_{i-1})+\beta(Y(t_i)-Y(t_{i-1}))
\\
\mbox{\hspace{3cm}}\displaystyle{}+\sigma\int_{t_{i-1}}^{t_i}\sqrt{\tau(s-)}dW(s)
+\rho(Z_\lambda(t_i)-Z_\lambda(t_{i-1})),
\end{array}
\end{equation}
where \begin{equation}\label{IV} Y(t)=\int_0^t \tau(s-)ds
\end{equation} is
the integrated trading volume/number of trades process. This
suggests introducing the discrete time quantities
\begin{equation}\label{defx}
X_i=X(t_i)-X(t_{i-1}),\quad Y_i=Y(t_i)-Y(t_{i-1}),\quad
Z_i=Z_\lambda(t_i)-Z_\lambda(t_{i-1})
\end{equation}
and
\begin{equation}
W_i=\frac1{\sqrt{Y_i}}\int_{t_{i-1}}^{t_i}\sqrt{\tau(s-)}dW(s).
\end{equation}
Furthermore, it is also convenient to introduce the discrete
quantity
\begin{equation}\label{defs}S_i=\frac{1}{\lambda}(Z_i-U_i).
\end{equation}
It is not difficult to
see (conditioning!) that $(W_i)_{i\geq1}$ is an iid $N(0,1)$
sequence independent from all other discrete quantities. We note
also that $(U_i,Z_i)_{i\geq1}$ is a bivariate iid sequence, but
$U_i$ and $Z_i$ are obviously dependent.

From now on, for notational simplicity, we consider the
equidistant grid with
\begin{equation}t_k=k\Delta,
\end{equation}
where $\Delta>0$ is fixed.
This implies
\begin{equation}\label{AR}\tau_i=\gamma \tau_{i-1}+U_i
\end{equation} and
\begin{equation}\label{defy}Y_i=\epsilon \tau_{i-1}+S_i,
\end{equation}where
\begin{equation}
\gamma=e^{-\lambda\Delta},\qquad\epsilon=\frac{1-\gamma}{\lambda}.
\end{equation}
Furthermore,
\begin{equation}\label{defxd}
X_i=\mu\Delta+\beta Y_i+\sigma\sqrt{Y_i}W_i+\rho Z_i.
\end{equation}
The sequence $(X_i,\tau_i)_{i\geq0}$ is clearly Markovian. From
now on we assume all moments of the stationary distribution of
$\tau_0$ exist.
\begin{assumption}\label{moments}
\begin{equation}
E[\tau_0^n]<\infty\qquad\forall n\in\mathbb N.
\end{equation}
\end{assumption}
In the estimating context we assume all moments are
finite with respect to all probability measures $P_\theta, \theta\in\Theta$ under
consideration, where~$\Theta$ is the parameter space.

No other assumptions are made, and all conditions required for
consistency and asymptotic normality of our estimator will be
proven rigorously from that assumption.
\begin{proposition}\label{Prop1}
We have for all $n\in\mathbb N$ that
\begin{equation}
E[Z_1^n]<\infty,\quad E[U_1^n]<\infty,\quad E[S_1^n]<\infty,
\end{equation}
and
\begin{equation}
E[Y_1^n]<\infty,\quad E[W_1^n]<\infty,\quad E[X_1^n]<\infty.
\end{equation}
Consequently the expectation of any (multivariate) polynomial in
$Z_1,U_1,S_1,\sqrt{Y_1},W_1,X_1$ exists under $P_\theta$.
\end{proposition}
\proof The proof is given in \cite[Proposition~$1$]{HP2007}.



Let us remark that, by the
stationarity, the above result holds also for
$Z_i,U_i,S_i,\sqrt{Y_i},W_i,X_i$ instead of
$Z_1,U_1,S_1,\sqrt{Y_1},W_1,X_1$, where $i\in\mathbb N$ is
arbitrary.
\section{A theoretical framework of the estimation procedure\label{main}}
\subsection{The estimating equations and their explicit solution}
The reader familiar with \cite{HP2007} will notice that the
following developments are quite similar to the paper mentioned,
the main (but important) difference is an additional estimating
equation for the new parameter $\sigma.$

For estimation purposes we consider a probability space on which a
parameterized family of probability measures is given:
\begin{equation}
\big(\Omega,\lijepof,\big\{P_\theta:\theta\in\Theta\big\}\big),
\end{equation}
where $\Theta=\{\theta\in\mathbb
R^6:\theta^1>0,\theta^2>0,\theta^3>0,\theta^6>0\}$. The data is
generated under the true probability measure $P_{\theta_0}$ with
some $\theta_0\in\Theta$. The expectation with respect to
$P_\theta$ is denoted  by $E_\theta[.]$ and with respect to
$P_{\theta_0}$ simply by $E[.]$.

We assume there is a process~$\mathbf{X}$ that is
BNS-DOU($\theta$) under~$P_\theta$. We want to find an estimator
for~$\theta_0$ using observations
$X_1,\ldots,X_n,\tau_1,\ldots,\tau_n$. We are interested in
asymptotics as $n\to \infty$. To that purpose let us consider the
following martingale estimating functions:
\begin{equation}\label{mef1}\renewcommand{\arraystretch}{1.5}
\begin{array}{ll}
G_n^1(\theta)=\sum_{k=1}^n\big[\tau_k       -f^1(\tau_{k-1},\theta)\big],&\qquad f^1(\iota,\theta)=E_\theta[\tau_1   |\tau_0=\iota]\\
G_n^2(\theta)=\sum_{k=1}^n\big[\tau_k\tau_{k-1}-f^2(\tau_{k-1},\theta)\big],&\qquad f^2(\iota,\theta)=E_\theta[\tau_1\tau_0|\tau_0=\iota]\\
G_n^3(\theta)=\sum_{k=1}^n\big[\tau_k^2     -f^3(\tau_{k-1},\theta)\big],&\qquad f^3(\iota,\theta)=E_\theta[\tau_1^2 |\tau_0=\iota]\\
G_n^4(\theta)=\sum_{k=1}^n\big[X_k       -f^4(\tau_{k-1},\theta)\big],&\qquad f^4(\iota,\theta)=E_\theta[X_1   |\tau_0=\iota]\\
G_n^5(\theta)=\sum_{k=1}^n\big[X_k\tau_{k-1}-f^5(\tau_{k-1},\theta)\big],&\qquad f^5(\iota,\theta)=E_\theta[X_1\tau_0|\tau_0=\iota]\\
G_n^6(\theta)=\sum_{k=1}^n\big[X_k\tau_k
-f^6(\tau_{k-1},\theta)\big],&\qquad
f^6(\iota,\theta)=E_\theta[X_1\tau_1|\tau_0=\iota]\\
G_n^7(\theta)=\sum_{k=1}^n\big[X_k^2
-f^7(\tau_{k-1},\theta)\big],&\qquad
f^7(\iota,\theta)=E_\theta[X_1^2 |\tau_0=\iota]
\end{array}
\end{equation}
\begin{lemma}
We have the explicit expressions
\begin{equation}
\begin{array}{l}\label{deff}
f^1(\iota,\theta)=\gamma\iota+(1-\gamma)\zeta\\
f^2(\iota,\theta)=\gamma \iota^2+(1-\gamma)\zeta \iota\\
f^3(\iota,\theta)=\iota ^2
\gamma^2+2\gamma(1-\gamma)\zeta\iota+(1-\gamma)^2\zeta^2+(1-\gamma^2)\eta\\
f^4(\iota,\theta)=\beta \epsilon \iota +\Delta \mu +\beta \Delta
\zeta-\beta \epsilon \zeta+\Delta \lambda
   \rho \zeta\\
f^5(\iota,\theta)=\beta \epsilon \iota^2 +(\Delta \mu +\beta
\Delta \zeta-\beta \epsilon \zeta+\Delta \lambda
   \rho \zeta)\iota\\
f^6(\iota,\theta)=\beta\epsilon\gamma \iota^2+\Delta\mu\gamma
\iota+\eta\lambda(\Delta\beta\epsilon\nu+\epsilon(2+\Delta\lambda\nu)\rho)+\zeta(\beta
\iota
(\epsilon(1-2\gamma)+\Delta\gamma)+\Delta\lambda(\mu\epsilon+\gamma
\iota \rho))\\
f^7(\iota,\theta)=\Delta^2 \eta  \nu  \beta^2+\epsilon^2 \eta \nu
\beta^2-2 \Delta \epsilon \eta  \nu \beta^2+\epsilon
   \iota (\epsilon (\iota-2 \zeta)+2 \Delta \zeta) \beta^2+4 \Delta \eta  \rho \beta-4
   \epsilon \eta  \rho \beta+2 \Delta \epsilon \mu  \iota \beta\\
\mbox{\hspace{1.5cm}}+2 \Delta^2 \mu
   \zeta
   \beta-2 \Delta \epsilon \mu  \zeta \beta+\Delta^2
\mu ^2+\Delta^2 \eta  \lambda^2 \nu
   \rho^2+\epsilon \sigma^2 (\iota-\zeta)+\Delta \sigma^2 \zeta\\
\mbox{\hspace{1.5cm}}+\lambda \left(2 \beta \eta
   \nu  \rho \Delta^2+2 \mu  \rho \zeta \Delta^2+2 \eta  \rho^2 \Delta-2 \beta \epsilon \eta
   \nu  \rho \Delta +2 \beta \epsilon \rho \iota \zeta
\Delta\right)+\displaystyle{\frac{2 \beta^2 \Delta
   \eta -4 \beta^2 \epsilon \eta }{\lambda}}\\
\mbox{\hspace{1.5cm}}+\displaystyle{\frac{\beta^2 \eta -\beta^2
\eta
   \gamma^2}{\lambda^2}}
\end{array}
\end{equation}
\end{lemma}
\proof  The formulas are special cases of the general moment
calculations given in~\cite[Appendix~A]{HP2007}.
\\
\hfill \qed

The estimator $\hat{\theta}_n$ is obtained by solving the
estimating equation $G_n(\theta)=0$ and it turns out that this
equation has a simple explicit solution.
\begin{proposition}\label{ESTIMATOR}
The estimating equation $G_n(\hat\theta_n)=0$ admits for every $n\geq2$
on the event
\begin{equation}
C_n=\big\{\xi_n^2-\xi_n^1\upsilon_n^1>0,\upsilon_n^2-(\upsilon_n^1)^2>0\big\}
\end{equation}
a unique solution
$\hat\theta_n=(\nu_n,\alpha_n,\lambda_n,\mu_n,\beta_n,\sigma_n,\rho_n)$
that is given by
\begin{equation}\label{thn}
\renewcommand{\arraystretch}{1.5}
\begin{array}{l}
\gamma_n=(\xi_n^2-\xi_n^1\upsilon_n^1)/(\upsilon_n^2-(\upsilon_n^1)^2);
\\
\zeta_n=\displaystyle\frac{\gamma_n\upsilon_n^1-\xi_n^1}{-1+\gamma_n}
\\
\eta_n
=-\displaystyle\frac{(-1+\gamma_n^2)(\upsilon_n^1)^2-\gamma_n^2\upsilon_n^2+\xi_n^3}{-1+\gamma_n^2}
\\
\lambda_n=-\log(\gamma_n)/\Delta;
\\
\epsilon_n=(1-\gamma_n)/\lambda_n;
\\
\beta_n=\displaystyle\frac{(\xi_n^5-\upsilon_n^1\xi_n^4)}{\epsilon_n(\upsilon_n^2-(\upsilon_n^1)^2)};
\\
\rho_n=\displaystyle\big(-\beta_n\epsilon_n(-(\upsilon_n^1)^2+\epsilon_n\lambda_n
(\eta_n+(\upsilon_n^1)^2-\upsilon_n^2)+\upsilon_n^2)-\xi_n^1\xi_n^4+\xi_n^6\big)/(2\epsilon_n\eta_n\lambda_n);
\\
\mu_n=\big(-\Delta\lambda_n\rho_n\zeta_n-\beta_n(\Delta\zeta_n+\epsilon_n(-\zeta_n+\upsilon_n^1))+\xi_n^4\big)/\Delta;\\
\sigma_n=\displaystyle\sqrt{a_n/b_n};\\
a_n=\frac{4\beta_n(-\Delta+\epsilon_n)\eta_n\lambda_n\rho_n+\beta_n^2(-2\Delta
\eta_n+\epsilon_n(\eta_n(2+\epsilon_n\lambda_n)+\epsilon_n\lambda_n((\upsilon_n^1)^2-\upsilon_n^2)))+
\lambda_n(-2\Delta\eta_n\lambda_n\rho_n^2-(\xi_n^4)^2+\xi_n^7)}{\lambda_n};\\
b_n=\Delta\zeta_n+ \epsilon_n(-\zeta_n+\upsilon_n^1);
\end{array}
\end{equation}
where
\begin{equation}\label{xin}
\renewcommand{\arraystretch}{2.0}
\begin{array}{llll}
\xi_n^1=\frac1n\sum\limits_{i=1}^n\tau_i &
\xi_n^2=\frac1n\sum\limits_{i=1}^n\tau_i\tau_{i-1} &
\xi_n^3=\frac1n\sum\limits_{i=1}^n\tau_i^2
\\
\xi_n^4=\frac1n\sum\limits_{i=1}^nX_i &
\xi_n^5=\frac1n\sum\limits_{i=1}^nX_i\tau_{i-1} &
\xi_n^6=\frac1n\sum\limits_{i=1}^nX_i\tau_i &
\xi_n^7=\frac1n\sum\limits_{i=1}^nX_i^2
\end{array}
\end{equation}
and
\begin{equation}\label{upn}
\begin{array}{ll}
\upsilon_n^1=\frac1n\sum\limits_{i=1}^n\tau_{i-1} &
\upsilon_n^2=\frac1n\sum\limits_{i=1}^n\tau_{i-1}^2
\end{array}
\end{equation}
\end{proposition}
\proof The first three equations $G_n^j(\theta)=0$, for $j=1,2,3$
contain only the unknowns $\zeta, \eta, \lambda$ and are easily
solved. In fact we get a familiar estimator for the first two
moments and the autocorrelation coefficient of an AR(1) process.
The last four equations $G_n^j(\theta)=0$, for $j=4,5,6,7$ can be
seen as a linear system for the unknowns $\mu,\beta,\rho,\sigma$,
once the other parameters have been determined. \\
\hfill\qed
\begin{remark}
The exceptional set $C_n$ could be simplified to
\begin{equation}
C_n'=\big\{\xi_n^2-\xi_n^1\upsilon_n^1>0\big\}
\end{equation}
Since the jump times and the jump sizes of the BDLP are
independent, and the former have an exponential distribution it
follows that $\tau_0,\ldots,\tau_n$ is with probability one not
constant, so $P[\upsilon_n^2-(\upsilon_n^1)^2>0]=1$. 
Furthermore, for finite $n$ we have
$P[\xi_n^2-\xi_n^1\upsilon_n^1\leq0]>0$. 
For definiteness we
put~$\hat\theta_n=0$ outside~$C_n$.
\end{remark}
\subsection{Consistency and asymptotic normality}
Let us investigate the consistency and asymptotic normality of the
estimator from the previous section.
\begin{theorem}\label{Thmmoments} We have $P(C_n)\to 1$ when $n\to\infty$ and the
estimator $\hat{\theta}_n$ is consistent on $\dis C_n$, namely
$$\hat{\theta}_n\asto\theta_0$$ on $C_n$
as\ $n\to \infty.$
\end{theorem}
\proof From \cite[Lemma~4]{HP2007} it follows that for all
integers $p,q,r\geq0$ we have
\begin{equation}\label{LLNHP07}
\frac1{n}\sum_{i=1}^nX_i^p\tau_i^q\tau_{i-1}^r\asto
E\big[X_1^p\tau_1^q\tau_0^r\big]
\end{equation}
as $n\to \infty$. Using this results it easily follows that
\begin{equation}
\xi_n^2-\xi_n^1\upsilon_n^1\to Cov(\tau_1,\tau_0)>0,
\end{equation}
so~$P(C_n)\to1$ as $n\to\infty$. Furthermore, from (\ref{LLNHP07})
it follows that the empirical moments in~(\ref{xin})
and~(\ref{upn}) converge to their theoretical counterparts,
$\xi_n^i\asto\xi^i$ and $\upsilon_n^i\asto\upsilon^i$, where
\begin{equation}
\begin{array}{l}
\xi^1=\zeta,\\
\xi^2=\zeta^2+\gamma\eta,\\
\xi^3=\zeta^2+\eta,\\
\xi^4=\Delta\big(\mu+(\beta+\lambda\rho)\zeta\big),\\
\xi^5=\Delta\zeta(\mu+\lambda\rho\zeta)+\beta(\epsilon\eta+\Delta\zeta^2),\\
\xi^6=2\epsilon\eta\lambda\rho+\Delta\zeta(\mu+\lambda\rho\zeta)+\beta(\epsilon\eta+\Delta\zeta^2),\\
\xi^7=\beta^2(2\Delta\eta-2\epsilon\eta+\Delta^2\lambda\zeta^2)/\lambda+
2\beta\big(2\Delta\eta\rho-2\epsilon\eta\rho\\
\qquad+\Delta^2\zeta(\mu+\lambda\rho\zeta)\big)
+\Delta\big(2\eta\lambda\rho^2+\sigma^2\zeta+\Delta(\mu+\lambda\rho\zeta)^2\big)
\end{array}
\qquad
\begin{array}{l}
\upsilon^1=\zeta,\\
\upsilon^2=\zeta^2+\eta.
\end{array}
\end{equation}
Plugging the limits into~(\ref{thn}) shows, after a short
mechanical calculation, that the estimator is consistent. \\
\hfill\qed\\
\hfill\\ \hfill
In order to prove asymptotic normality, 
we use the general framework and results of \cite{SO99}. 
We extend the theory in the case of a bivariate Markov process. To
apply \cite[Theorem~2.8]{SO99}, requires to show that
\cite[Condition~2.6]{SO99} is satisfied.

\begin{proposition}\label{prop27}The Condition $2.6$ of \cite{SO99} is
satisfied.
\end{proposition}
\proof  For a concise vector notation we introduce
\begin{equation}\label{Xik}
\Xi_k=(\tau_k,\tau_k\tau_{k-1},\tau_k^2,X_k,X_k\tau_{k-1},X_k\tau_k,X_k^2)^\top,
\end{equation}and we write the estimating equations in the form
\begin{equation}
G_n^i(\theta)=\sum_{k=1}^n\big[\Xi_k^i-f^i(\tau_{k-1},\theta)\big],
\qquad i=1,\ldots,7.
\end{equation}
Looking at (\ref{deff}) we note that $f^i(\iota,\theta)$ is a
polynomial in $\iota,$ namely
\begin{equation}\label{fpol}f^i(\iota,\theta)=\sum_{k=0}^{p_i}\phi_{i,k}(\theta)\iota^k,\end{equation}
where the degree $p_i$ and the coefficients $\phi_{i,k}(\theta),$
which  are smooth functions in $\theta,$ can be read off
from~(\ref{deff}).
Now
the proof is completely analogous to that of
\cite[Proposition~4]{HP2007}. \\
\hfill\qed

Finally, we have all the ingredients for proving the following result.
\begin{theorem}
The estimator
\begin{equation}
\hat{\theta}=(\nu_n,\alpha_n,\lambda_n,\mu_n,\beta_n,\sigma_n,\rho_n)
\end{equation}
is asymptotically normal, namely
\begin{equation}\label{CLTtheta}
\sqrt{n}\big[\hat{\theta}_n-\theta_0\big]\inlawto N(0,T),
\end{equation}
as $n\to\infty$, where
\begin{equation}
T=A(\theta)^{-1}\Upsilon\big(A(\theta)^{-1}\big)^T,\qquad
\Upsilon_{ij}=E[Cov(\Xi_1^i,\Xi_1^j|\tau_0)]
\end{equation}
and
\begin{equation}A_{i,j}(\theta)=E\bigg[\frac{\partial}{\partial\theta_j}f^i(\tau_0,\theta)\bigg].
\end{equation}
\end{theorem}
\begin{remark}Looking at~(\ref{fpol}), we see that
$\dis\frac{\partial}{\partial\theta_j}f^i(\tau_0,\theta)$ is a
polynomial in $\tau_0$ and thus we can find explicit expressions
for the entries of $A.$ Similar arguments allow us to obtain
explicit expressions for $\Upsilon,$ see \cite{HP2007}.
\end{remark}
\proof
From \cite[Proposition~3]{HP2007} it follows that
\begin{equation}\label{CW}
\frac{1}{\sqrt{n}}G_n(\theta_0)\inlawto N(0,\Upsilon),
\end{equation}
as $n\to\infty$, where
\begin{equation}
\Upsilon_{ij}=E\big[Cov(\Xi_1^i,\Xi_1^j|\tau_0)\big]
\end{equation}and $\dis\big(\Xi^i,~i=1,\ldots,7\big)$ is defined by (\ref{Xik}).
Using the just obtained result and  Proposition~\ref{prop27}, the
result follows directly from \cite[Theorem2.8]{SO99}.\\
\hfill\qed
\section{A numerical illustration of the finite sample performance of the estimator\label{numerical}}
\subsection{Description of the model and its parameter values}

To illustrate the results from the previous sections numerically,
we consider the $\Gamma$-OU model from Section~\ref{Sec-GaOU},
where the trading volume $\tau$ has a stationary gamma
distribution. The corresponding BDLP $Z$ is a compound Poisson
process with intensity $\nu$ and jumps from the exponential
distribution with parameter $\alpha.$ We use as time unit one year
consisting of 250 trading days. The true parameters are

{\footnotesize \begin{equation} \nu=6.17,\quad \alpha=1.42,\quad
\lambda=177.95,\quad \beta=-0.015,\quad \rho=-0.00056,\quad
\mu=0.435,\quad \sigma=0.087.
\end{equation}}

The parameters imply that there are on average $4.4$ jumps per day
and the jumps in the BDLP and in the trading volume are
exponentially distributed with mean and standard deviation
$0.704$. The interpretation is, that typically every day $4$ or
$5$ new pieces of information arrive and make the trading volume
process jump. The stationary mean of the trading volume is $4.35,$
and of the variance is $0.033$. Hence, if we define instantaneous
volatility to be the square root of the variance, it will
fluctuate around $18\%$ in our example. The half-life of the
autocorrelation of the variance process is about a day.

In our example annual log returns have (unconditional) mean
$-6.5\%$ and annual volatility $18.2\%$. We will perform the
estimation procedure for two different sample sizes, namely $2500$
and $8000,$ corresponding to $10$ years and $32$ years
respectively, with $250$ daily observation per year.

\subsection{Simulation study}

We first simulate 1000 samples of $n=2500$ equidistant
observations of $X_i$ and $\tau_i,~i=1,\ldots,n.$ Table~$1$
summarizes the estimation results of our simulation study
concerning the parameters $\nu,
\alpha,\lambda,\mu,\beta,\sigma,\rho.$

\begin{figure}[h]
\begin{center}
\includegraphics[width=14cm,height=4.5cm]{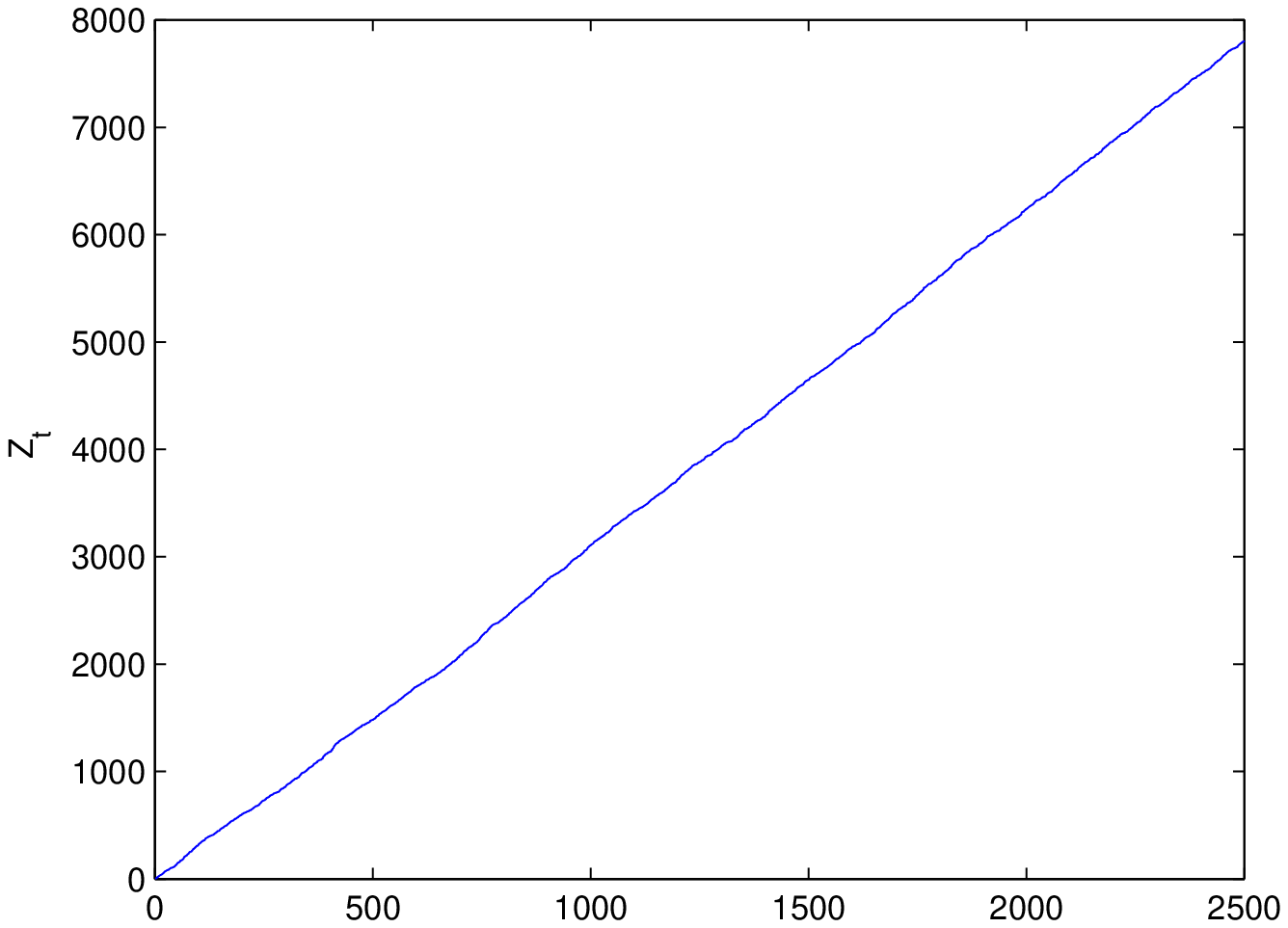}\\
\includegraphics[width=14cm,height=4.5cm]{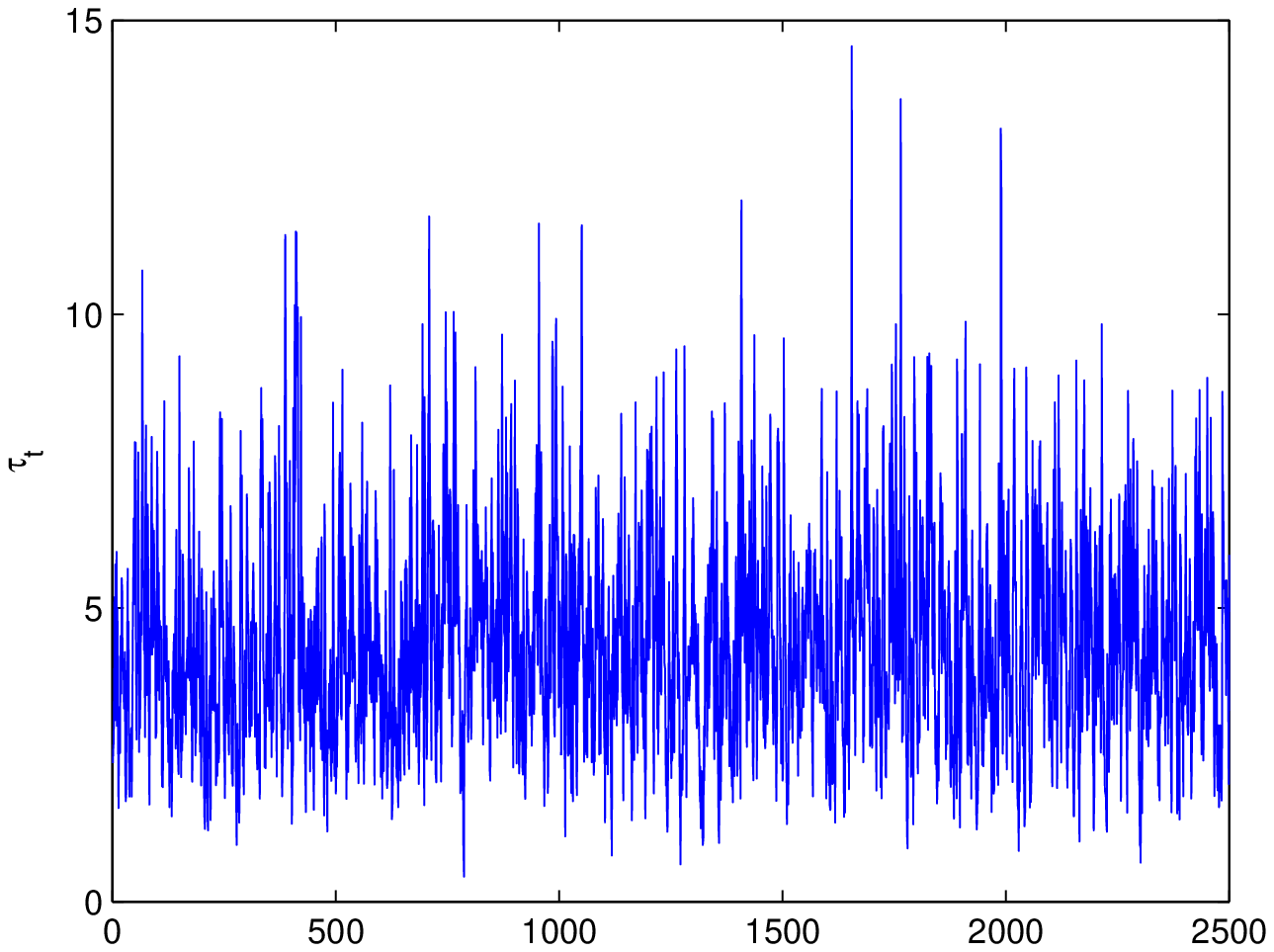}\\
\includegraphics[width=14cm,height=4.5cm]{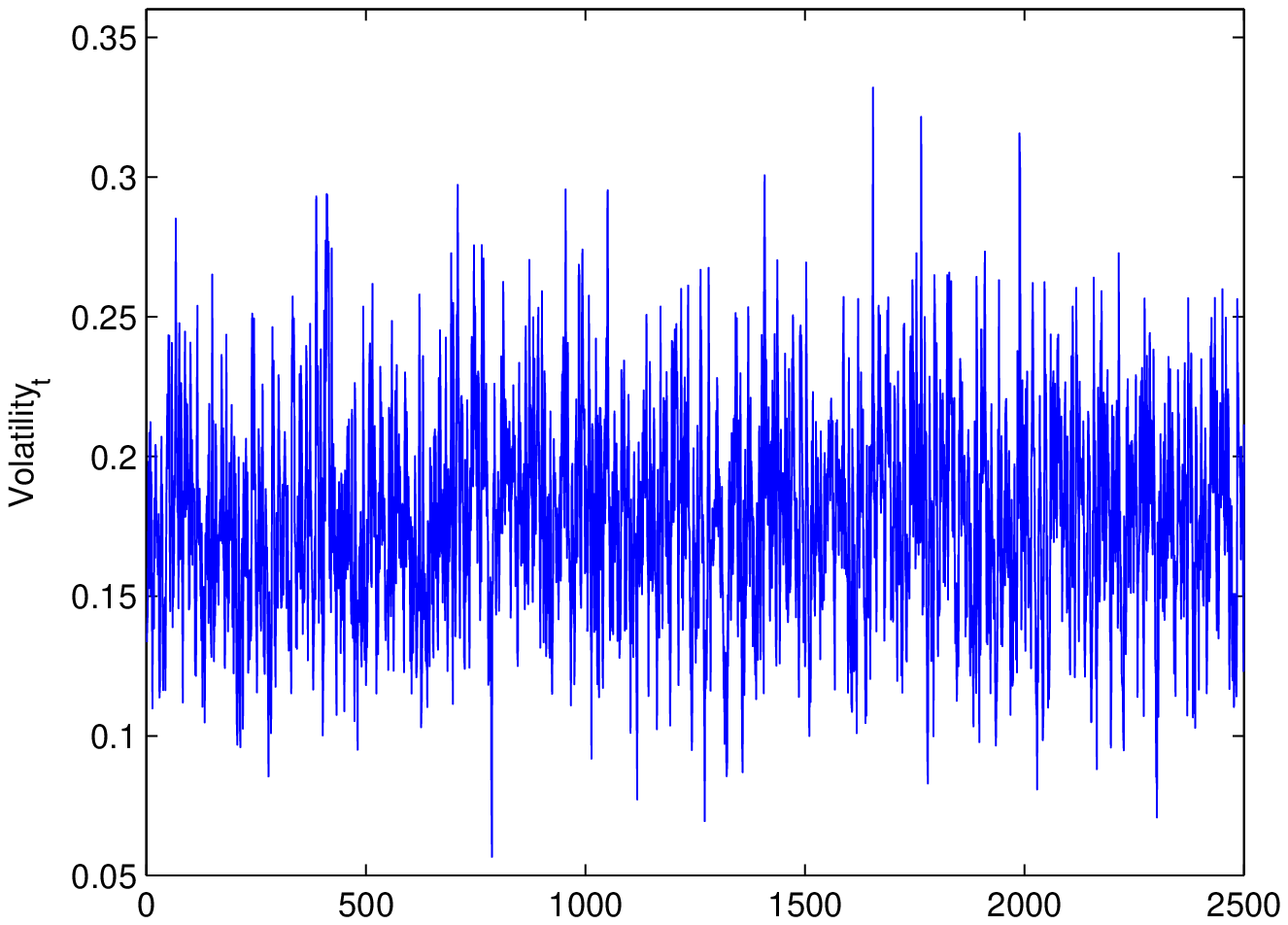}\\
\includegraphics[width=14cm,height=4.5cm]{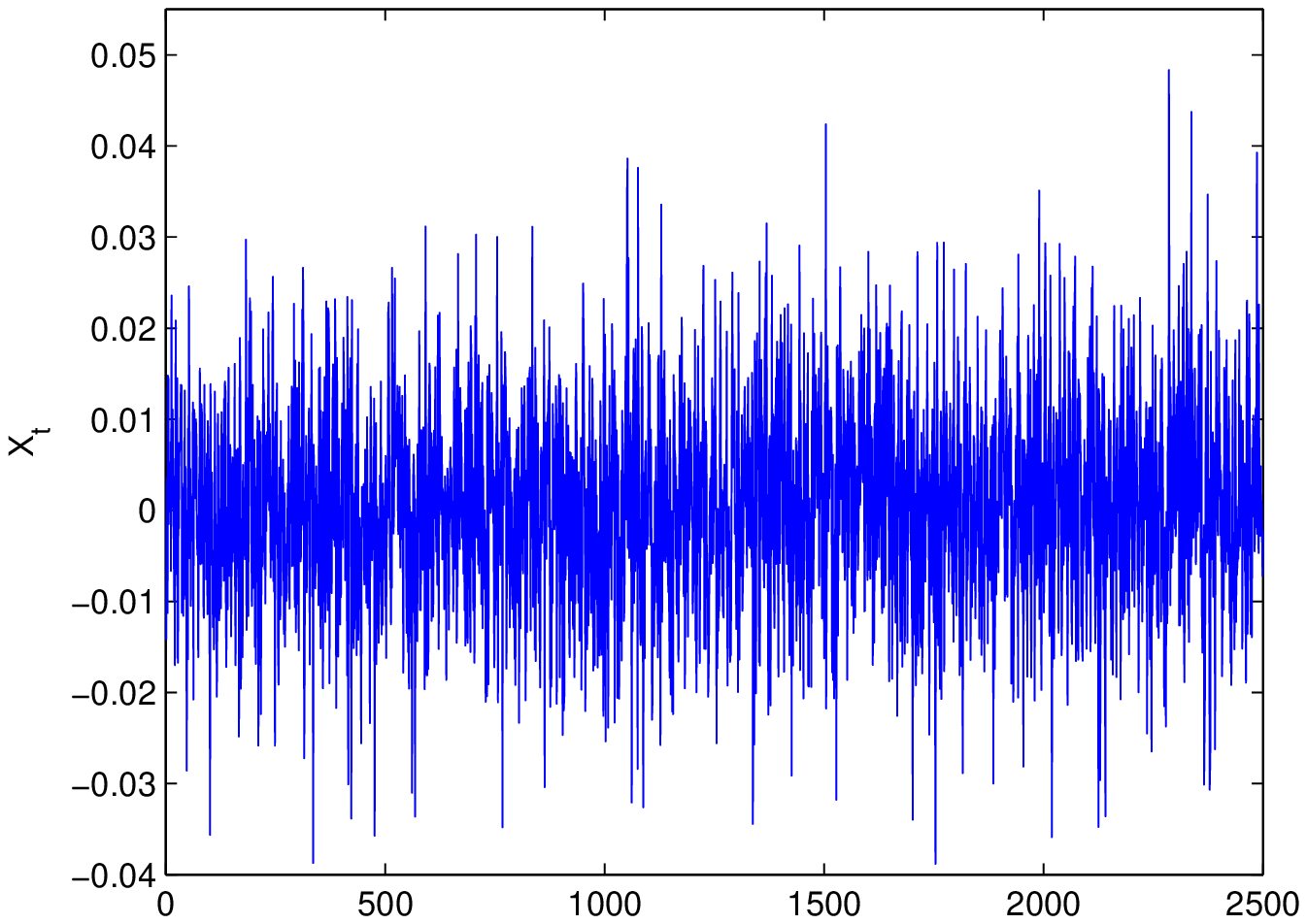}\\
\end{center}
\caption{Instantaneous BDLP $Z,$ number of trades/trading volume
$\tau_i,$ the volatility process $\sigma\times\sqrt{\tau_i}$ and
daily log returns $X_i$.\label{path}}
\end{figure}

Figure~\ref{path} displays a simulation of ten years of daily
observations from the background driving \Levy{} process,  the
instantaneous trading volume process, the volatility process and
log returns for $i=1,\ldots,2500$.
The empirical mean of all the estimated parameter values
$\nu_n,\alpha_n,\lambda_n,\mu_n,\beta_n,\sigma_n,\rho_n$ is shown
in the first line, with the empirical standard deviations in
brackets. We also estimated mean square error (MSE) and mean
absolute error (MAE), again with the standard deviation in
brackets. The corresponding results for a sample size of $n=8000$
observations are reported in the last three lines of
Table~\ref{Table3} and Table~\ref{Table4}.


\begin{table}
\begin{center}
\begin{tabular}{|c|c|c|c|}
  \hline $n=2500$ & $\nu_n$ & $\alpha_n$ & $\lambda_n$  \\
  \hline
  Mean & 6.2145 (0.2552)& 1.435 (0.0588) & 177.865 (8.9257)\\
  MSE & 0.0672 (0.1046) & 0.0036 (0.0055) & 79.5956 (115.6766) \\
  MAE & 0.2016 (0.1629) &0.047 (0.0369) & 7.0692 (5.4454) \\
  \hline $n=8000$ & $\nu_n$ & $\alpha_n$ & $\lambda_n$  \\
\hline
 Mean &6.1642 (0.1424) &1.4186 (0.0329) & 177.1342 (5.208) \\
  MSE &0.0203 (0.0283) & 0.0011 (0.0016)& 27.7663 (39.1414) \\
  MAE & 0.1135 (0.0862)& 0.0259 (0.0203)& 4.2191 (3.1584)\\
  \hline
\end{tabular}
\caption{Estimated means, MSE and MAE for the parameters
$\nu_n,\alpha_n,\lambda_n$ and the corresponding standard
deviations in brackets. The true values are
$\nu=6.17,\alpha=1.42,\lambda=177.95.$\label{Table3}}
\end{center}
\end{table}
\begin{table}
\begin{center}
\begin{tabular}{|c|c|c|c|c|}
  \hline $n=2500$ & $\mu_n$ & $\beta_n$ & $\sigma_n$ & $\rho_n$ \\
  \hline
  Mean & 0.4402 (0.1849)& 0.0148 (0.053) & 0.0871 (0.0013)& {\footnotesize $-5.65\cdot 10^{-4} ~(1.43\cdot 10^{-4})$}\\
  MSE &  0.0342 (0.0492)& 0.0028 (0.0039) & {\footnotesize $1.82\cdot 10^{-6} ~(2.33\cdot 10^{-6})$} & {\footnotesize $2\cdot 10^{-8}~ (3\cdot 10^{-8})$} \\
  MAE &  0.1483 (0.1105) & 0.0428 (0.0313) & 0.0011 (0.0008) & {\footnotesize $1.12\cdot 10^{-4} ~(8.85\cdot 10^{-5})$}\\
  \hline $n=8000$ & $\mu_n$ & $\beta_n$ & $\sigma_n$ & $\rho_n$ \\
\hline
 Mean & 0.4388 (0.1002) & 0.0129 (0.0278)& {\footnotesize $ 0.0872~ (7.45\cdot 10^{-4})$}& {\footnotesize $-5.55\cdot 10^{-4}~ (8.07\cdot 10^{-5})$}\\
  MSE & 0.01 (0.0138) & 0.0008 (0.0011)& {\footnotesize $5.56\cdot 10^{-7} ~(8.35\cdot 10^{-7})$}& {\footnotesize $6.5 \cdot 10^{-9} ~(1\cdot 10^{-8})$}\\
  MAE & 0.08 (0.0604) & 0.0222 (0.0169)&{\footnotesize $5.81\cdot 10^{-4} ~(4.68\cdot 10^{-4})$} & {\footnotesize $6.3\cdot 10^{-5} ~(5.04\cdot 10^{-5})$} \\
  \hline
\end{tabular}
\caption{Estimated means, MSE and MAE for the parameters
$\mu_n,\beta_n,\sigma_n,\rho_n$ and the corresponding standard
deviations in brackets. The true values are $\beta=-0.015,
\rho=-0.00056,\mu=0.435,\sigma=0.087.$\label{Table4}}
\end{center}
\end{table}

When one compares the estimates for the different sample sizes, it
can be seen that the MSE reduces for all seven estimators, when
the sample size is increased and the reduction is roughly of a
factor of $4$ which would correspond to the asymptotic properties
of the estimators.
\subsection{The asymptotic covariance matrix and the finite sample distribution of the estimator}

As our goal is an analysis of the estimator, we do not estimate
the asymptotic covariance, but evaluate the explicit expression
using the true parameters. Denoting the vector of asymptotic
standard deviations of the estimates and the correlation matrix by
$s/\sqrt{n}$ resp.~$r$ we have
\begin{equation}
s=[12.0257,~ 2.7878,~ 443.85,~ 9.0211,~ 2.5536,~0.0657,~
0.007]^{T}
\end{equation}

{\footnotesize \begin{equation}r=\left[
\begin{array}{ccccccc}
1 & 0.938 & 0.5778 & 0.0074 & 0.0511 & 0.0062 & -0.0026\\
  0.938 & 1 & 0.5738 & 0.0076 & 0.0507 & 0.0126 & -0.0039\\
  0.5778 & 0.5738 & 1 & 0.011 & 0.0884 & -0.00056 & -6.2\cdot
  10^{-17}\\
  0.0074 &  0.0076 & 0.011 &  1 & -0.8265 & -0.0128 & 0.0296\\
  0.0511 & 0.0507 &  0.0884 & -0.8265 &  1 & 0.012 & -0.5148 \\
  0.0062 &  0.0126 & -0.00056 & -0.0128 &  0.012 &  1 & -0.0045\\
 -0.0026 & -0.0039 & -6.2\cdot
  10^{-17} &  0.0296 & -0.5148 & -0.0045 & 1
\end{array}
\right]
\end{equation}}

We will discuss what this values of $s$ implies for the sample
size of $2500$ below. The correlations among parameters related to
the returns distribution, namely $\mu, \sigma, \rho$ and $\beta,$
are rather small except for $\beta$ and $\rho.$ In contrast to
that, correlations among the trading volume parameters, namely
$\nu, \alpha$ and $\lambda$ are very high. Theoretically, this can
be addressed using the optimal martingale estimating function
approach, even though the corresponding equations can not be
solved explicitly and the optimal estimator has to be obtained by
numerical optimization, see \cite{HP2006} for developments in this
direction.

\begin{figure}[h]
$$
\begin{array}{cc}
\includegraphics[width=7cm,height=4cm]{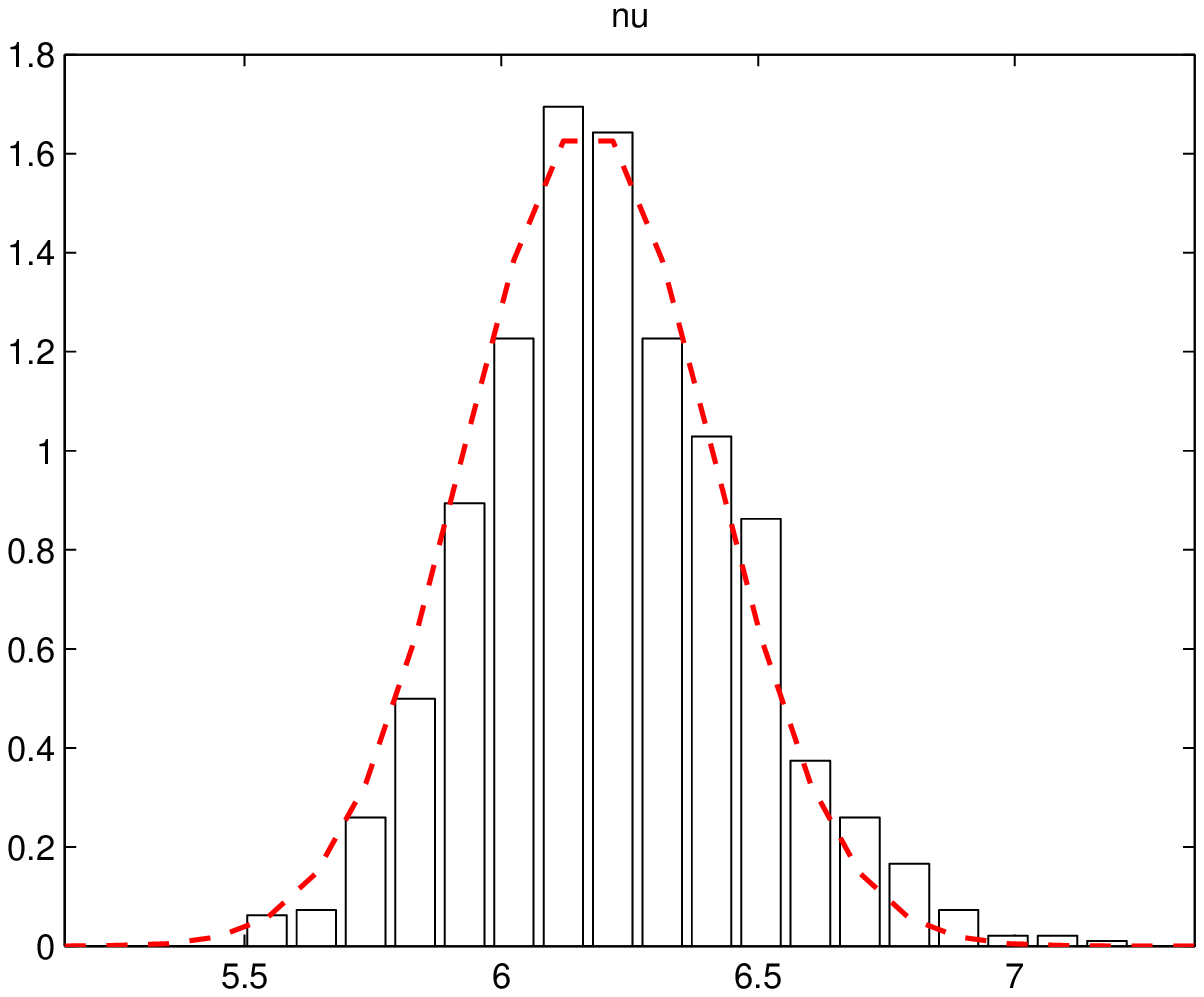} & \includegraphics[width=7cm,height=4cm]{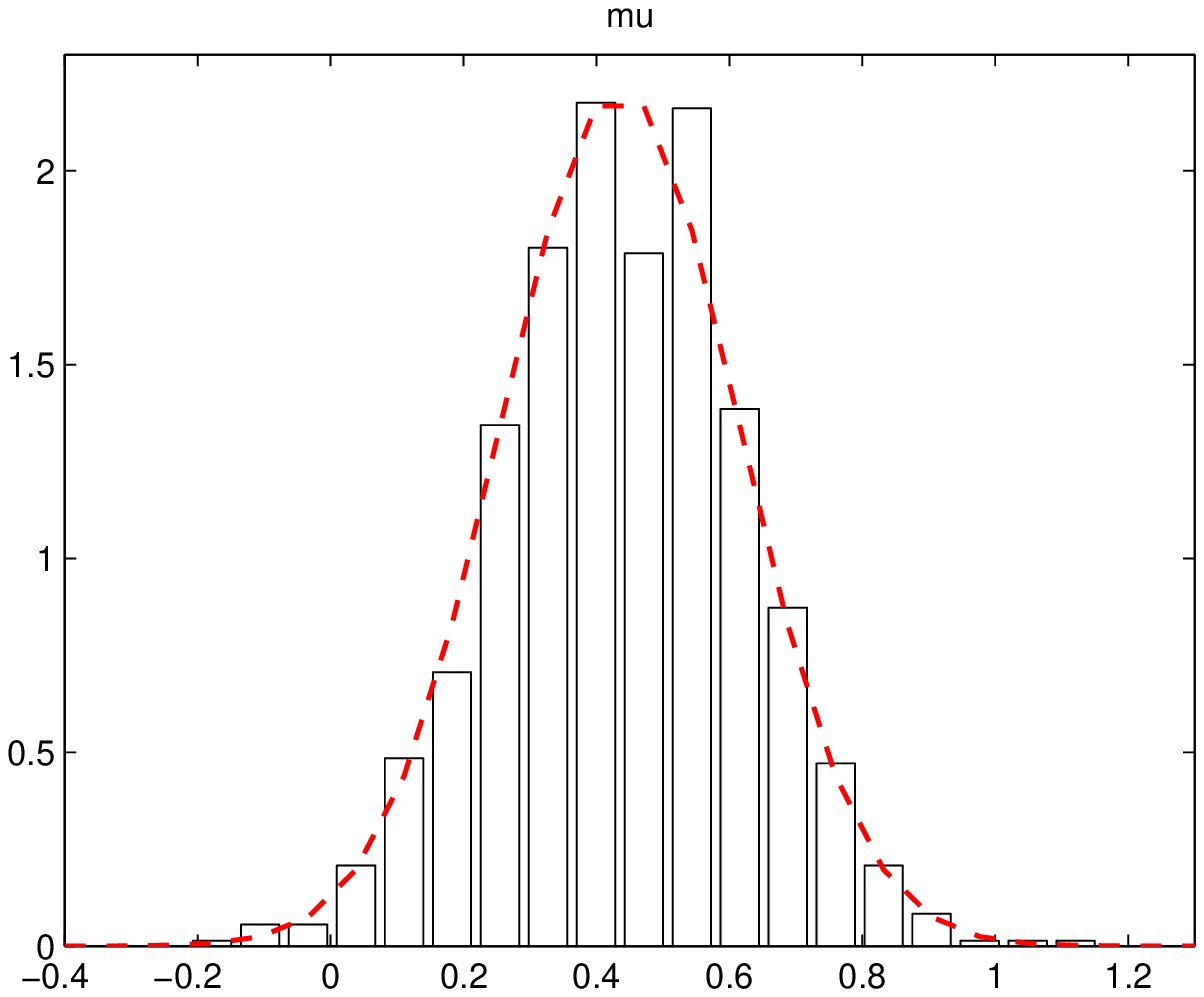}\\  
\includegraphics[width=7cm,height=4cm]{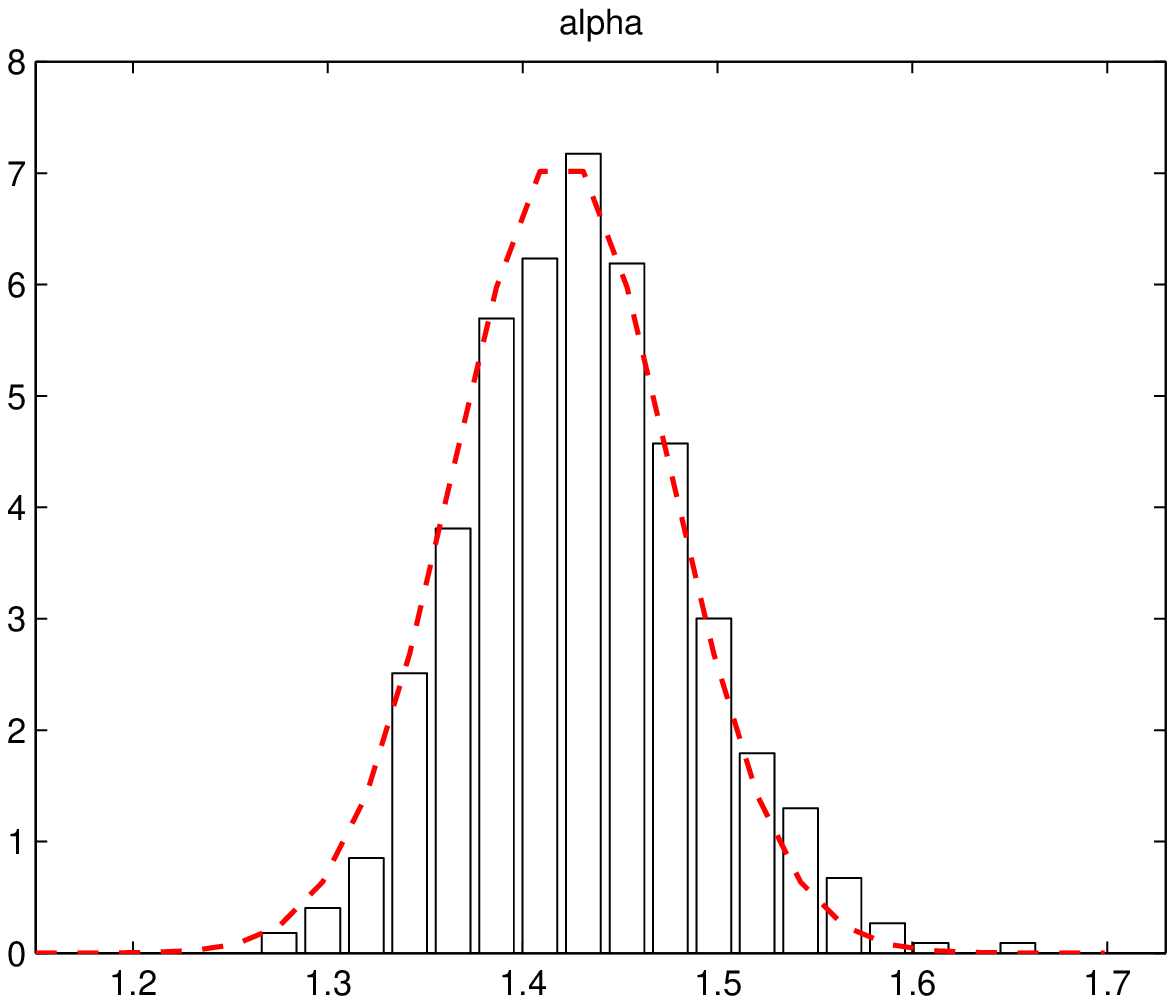} & \includegraphics[width=7cm,height=4cm]{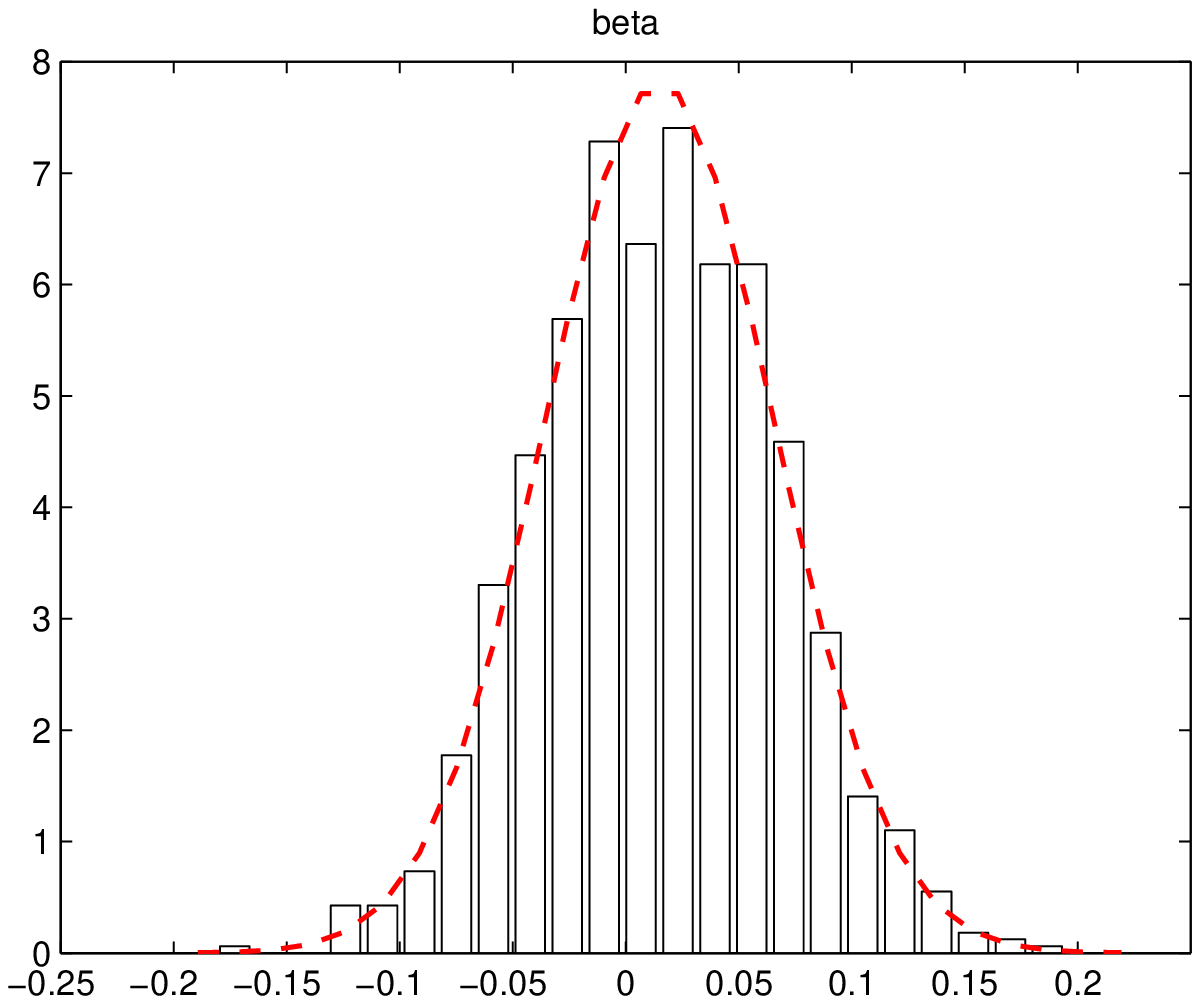}\\  
\includegraphics[width=7cm,height=4cm]{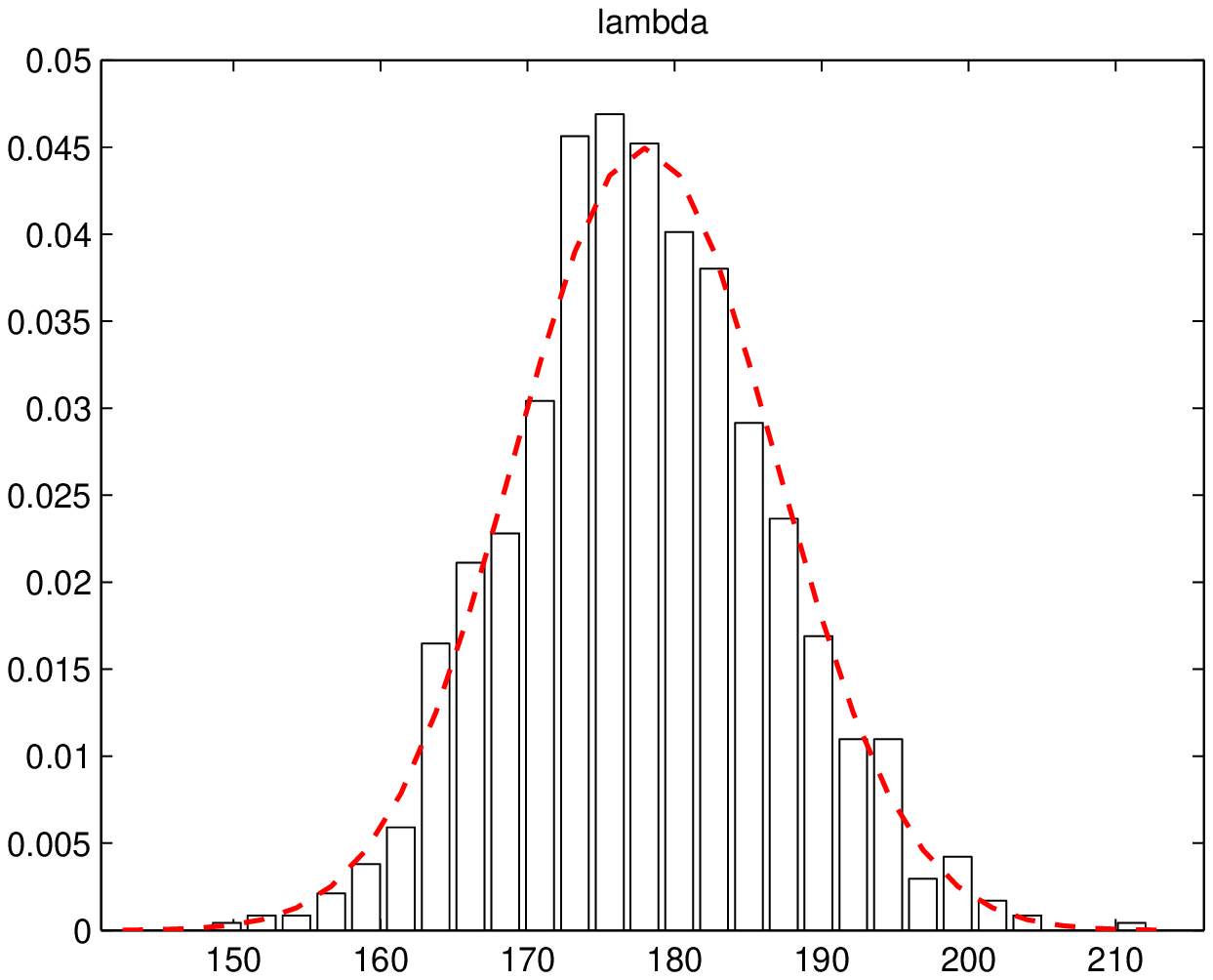} & \includegraphics[width=7cm,height=4cm]{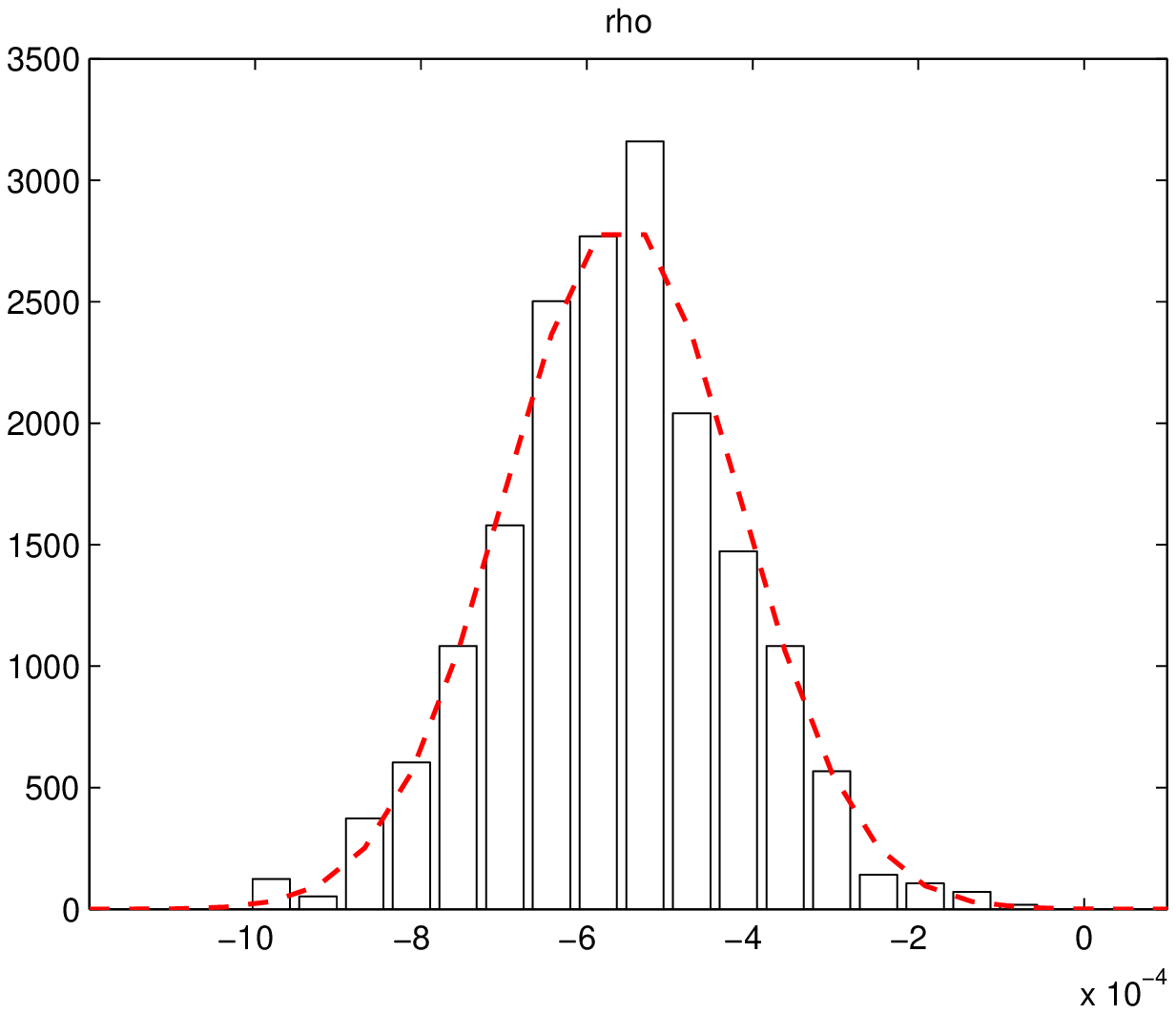}\\
\includegraphics[width=7cm,height=4cm]{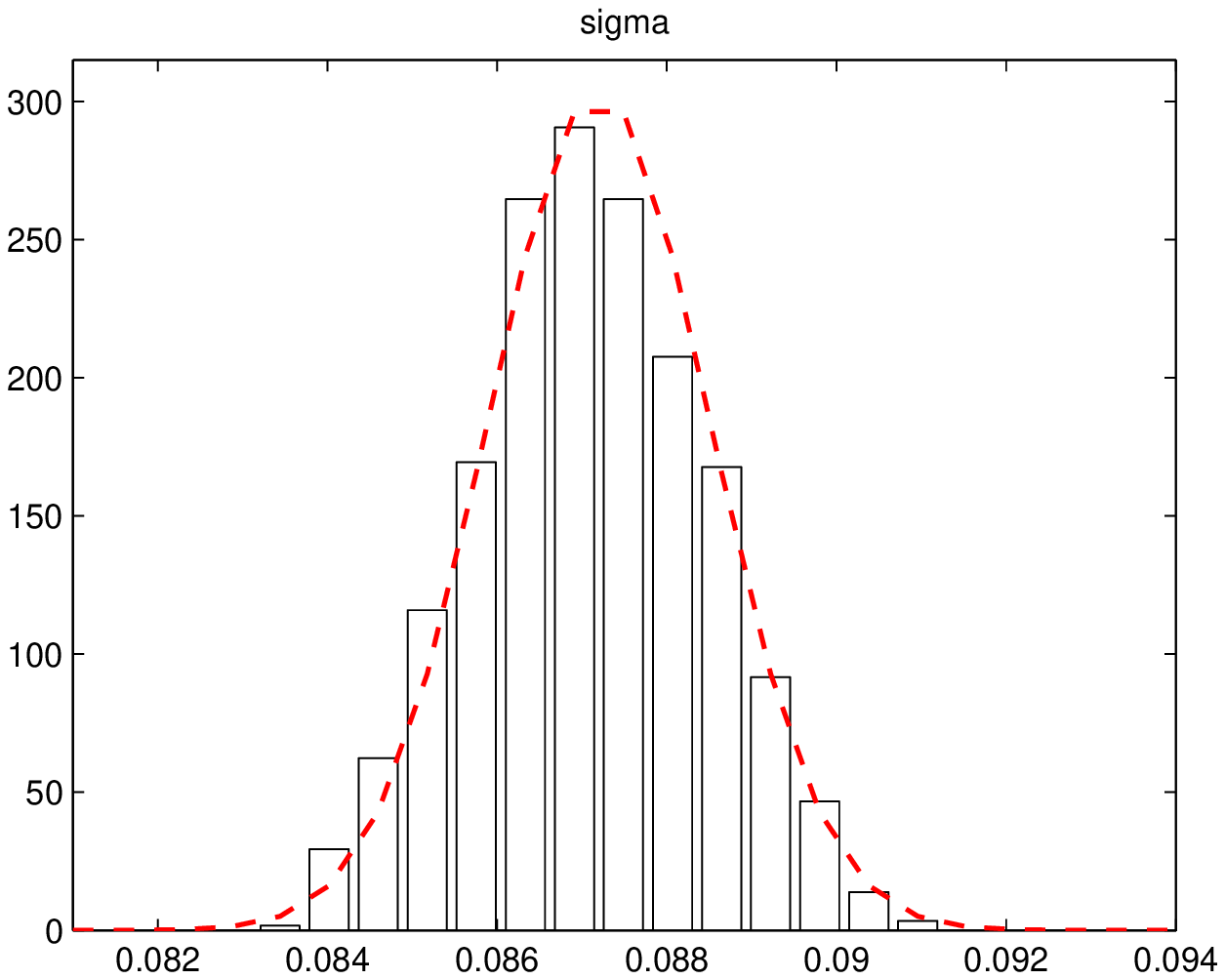}
\end{array}
$$
\caption{Simulated distribution of the simple estimators for the
$\Gamma$-OU model. The histograms are produced from $m=1000$
replications consisting of $n=2500$ observations each. The true
values are $\nu=6.17$, $\alpha=1.42$, $\lambda=177.95$,
$\mu=0.435$, $\beta=-0.015$, $\rho=-0.00056$, $\sigma=0.087$. The
standard deviations used for the normal curves are taken from the
explicit asymptotic results, not estimated.\label{his}}
\end{figure}

Figure~\ref{his} illustrates the empirical distribution of the
simple estimators for the $\Gamma$-OU model. The histograms are
produced from $m=1000$ replications consisting of $n=2500$
observations each, corresponding to 10 years with 250 daily
observations per year. Both from the graphs and the asymptotic
standard deviations we see that the parameter $\nu$ can be
estimated quite accurately to at least one digit of precision. The
parameter $\alpha$ is estimated even better with almost two digits
of precision. The autocorrelation parameter $\lambda$ is estimated
slightly less accurate. The parameter $\sigma$ which connects the
trading volume/number of trades and volatility is estimated quite
accurate with one to two digits of precision. This means that if
the relation between volatility and trading volume is exploited,
not too much uncertainty is introduced by the estimating
procedure. This can be also very promising for option pricing
purposes. The bad quality of the estimator for $\beta$ is neither
surprising nor very troublesome. It has little impact on the
model. The main reason for including the parameter $\beta$ in the
specification of BNS models is, for derivatives pricing: A
risk-neutral BNS-model must have $\beta=-1/2$. In most
applications working under a physical probability measure
$\beta=0$ can be assumed without much loss of generality or
flexibility. For the same reason the parameter $\mu$ is not very
relevant although it can be estimated more accurate than $\beta.$
Even though the value of the leverage parameter $\rho$ is rather
small, it can be estimated very accurately.
\subsection{Estimation of the volatility $\sqrt{V_t}$}

\begin{figure}[h]
\includegraphics[width=14cm,height=7cm]{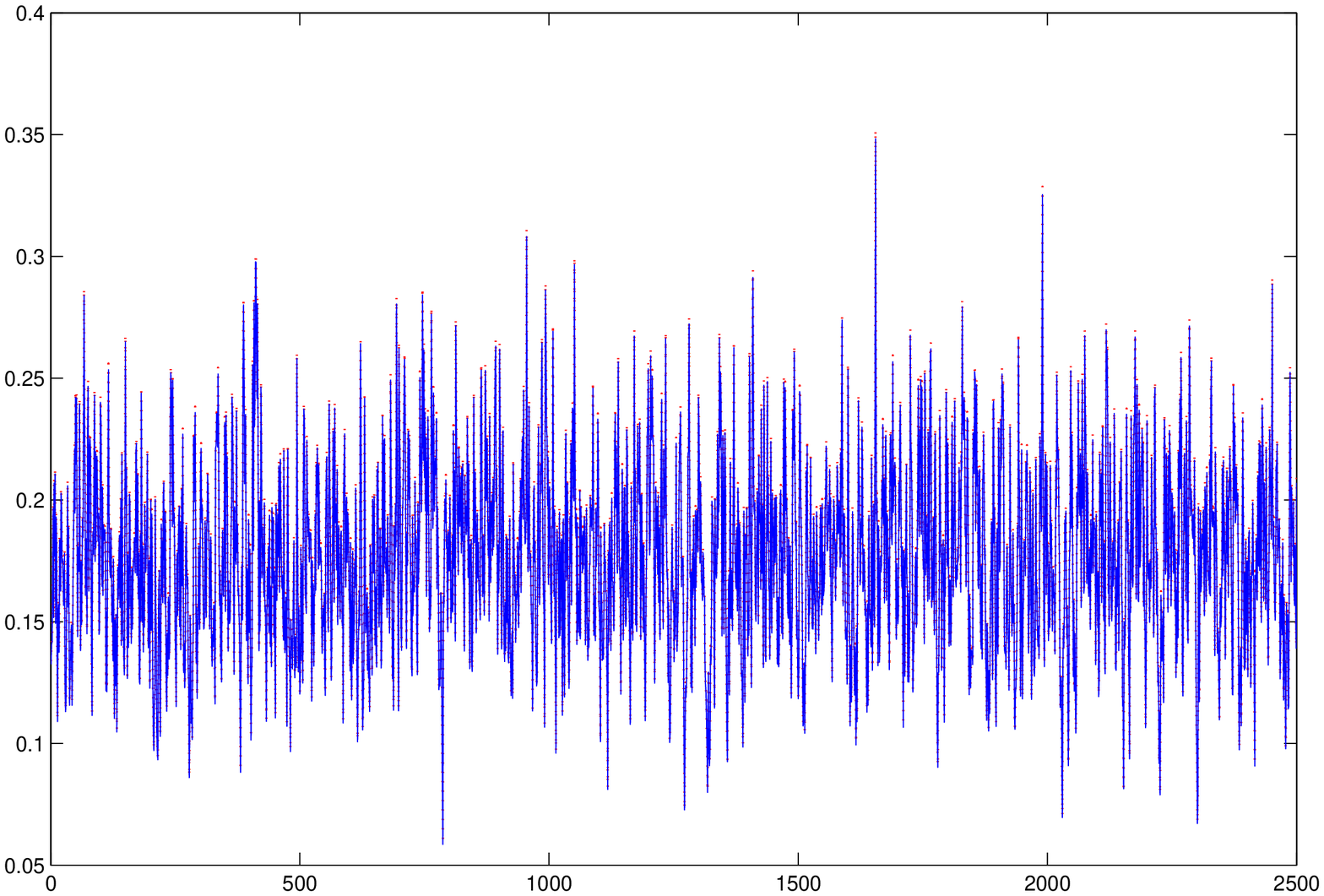}\\
\caption{Sample paths of $\sigma\sqrt{\tau_t}$ (solid line) and
$\displaystyle\hat{\sigma}\sqrt{\tau_t}$ (dashed line) of one
simulation.\label{Vlt_plot}}
\end{figure}

Recall from (\ref{trades}), that we assumed that the instantaneous
variance is a multiple of trading volume and thus we have the
following equation for the volatility
\begin{equation}\label{Volatility}\sqrt{V_i}=\sigma\sqrt{\tau_i},\qquad i\in\mathbb{N}.
\end{equation}
Since we observe $\tau_i$ at integer times, an estimate of the
volatility process $\hat{\sigma}\sqrt{\tau_i}$ can therefore be
calculated from (\ref{Volatility}) and it is plotted in Figure
\ref{Vlt_plot} together with the exact volatility
$(\sigma\sqrt{\tau_i})_{i\geq 1}$ for one simulated path. The
estimator for $\sigma$ is calculated from the simulated path and
since $\hat{\sigma}$ is very accurate, the two graphs are almost
indistinguishable.

Next we investigate the goodness of fit of our estimation method
by a residual analysis. Recall from (\ref{defxd}), the estimated
residuals are given by
\begin{equation}\label{residuals}\big(X_i-\hat{\mu}\Delta-\hat{\beta}Y_i-\hat{\rho}Z_i\big)/\hat{\sigma}\sqrt{Y_i},\qquad
i\in\mathbb{N}
\end{equation}where the integrated instantaneous trading volume $Y_i$ is given by (\ref{defy}).
The quantity $Z_i$ is not observable, but we can find an
approximation $\dis \hat{Z}_i$ as follows. For the integral we use
simple Euler approximations
\begin{equation}Y_i=\int_{t_{i-1}}^{t_i}\tau(s-)ds\approx\tau_i\Delta\quad\mathrm{and}\quad
\int_{t_{i-1}}^{t_i}\sqrt{\tau(s)}dW(s)\approx\sqrt{\tau_i\cdot\Delta}~\varepsilon_i,\qquad
i\in\mathbb{N}
\end{equation}
with $\varepsilon_i$ are approximately $N(0,1)$ and i.i.d. The
estimated residuals are given
by  
\begin{equation}\label{returnsap}
\hat{\varepsilon}_i=\big(X_i-\hat{\mu}\Delta-\hat{\beta}\tau_i\Delta-\hat{\rho}\hat{Z}_i\big)/\hat{\sigma}\sqrt{\tau_i\cdot\Delta}\nonumber\\
\end{equation}where
\begin{equation} \hat{Z}_i= (\hat{\lambda}\Delta+1)\tau_i-\tau_{i-1}.
\end{equation}
Since we assume that $(W_i)_{i\geq 0}$ is an iid $N(0,1)$
sequence, our goal is to check if the residuals
$\hat{\varepsilon}(\cdot)$ are iid and $N(0,1).$ The residuals
should be symmetric around zero and thus their mean and skewness
should be close to zero. Furthermore, we expect the kurtosis to be
close to three. Consequently, we estimated mean, MSE, MAE and the
corresponding standard deviations for the mean, the standard
deviation, the skewness and the kurtosis of the residuals
$\hat{\varepsilon}_i$ based on $1000$ simulations. The results for
both sample sizes are reported in Table \ref{Table5} and indicate
a reasonable fit.
\begin{table}
\begin{center}
\begin{tabular}{|c|c|c|c|c|}
  \hline $n=2500$ & mean$(\hat{\varepsilon})$ & Std$(\hat{\varepsilon})$ & skew$(\hat{\varepsilon})$ & kurt$(\hat{\varepsilon})$ \\
  \hline
  Mean & 0.11753 (0.03455)& 1.02865 (0.0067) & -0.04114 (0.06107)& 3.3018 (0.17843)\\
  MSE &  0.015(0.0086)& 0.00087 (0.00039) & 0.00542 (0.00753) & 0.12289 (0.14759) \\
  MAE &  0.11753 (0.03455) & 0.02865 (0.0067) & 0.05885 (0.04423) & 0.30291 (0.17654)\\
  \hline $n=8000$ & mean$(\hat{\varepsilon})$ & Std$(\hat{\varepsilon})$& skew$(\hat{\varepsilon})$ & kurt$(\hat{\varepsilon})$ \\
\hline
 Mean & 0.1166 (0.01951) & 1.02838 (0.00373)& -0.03876 (0.03476) & 3.29937 (0.10449)\\
  MSE & 0.01398 (0.00456) & 0.00082 (0.00021)& 0.00271 (0.00333) & 0.10053 (0.07529) \\
  MAE & 0.1166 (0.01951) & 0.02838 (0.00373)& 0.04335 (0.02883) & 0.299938 (0.10446) \\
  \hline
\end{tabular}
\caption{Estimated mean, MSE and MAE for the mean, standard
deviation, skewness and kurtosis of the residuals with
corresponding estimated standard deviations in
brackets.\label{Table5}}
\end{center}
\end{table}
The correlation of the squared residuals was checked by performing
a Ljung-Box test for each sample. For $n=2500$ we computed the
test statistic based on $50=\sqrt{2500}$ lags and had to reject
the null hypothesis of no correlation $60$ times out of 1000
simulations at the $0.05$ level. Whereas for $n=8000$ the test
statistic was computed using $90\approx\sqrt{8000}$ lags and the
null hypothesis was rejected $66$ times out of $1000$ simulations
again at the $0.05$ level.
\section{Real data analysis\label{empirical}}
\begin{figure}[h]
$$
\begin{array}{cc}
\includegraphics[width=7.5cm,height=7cm]{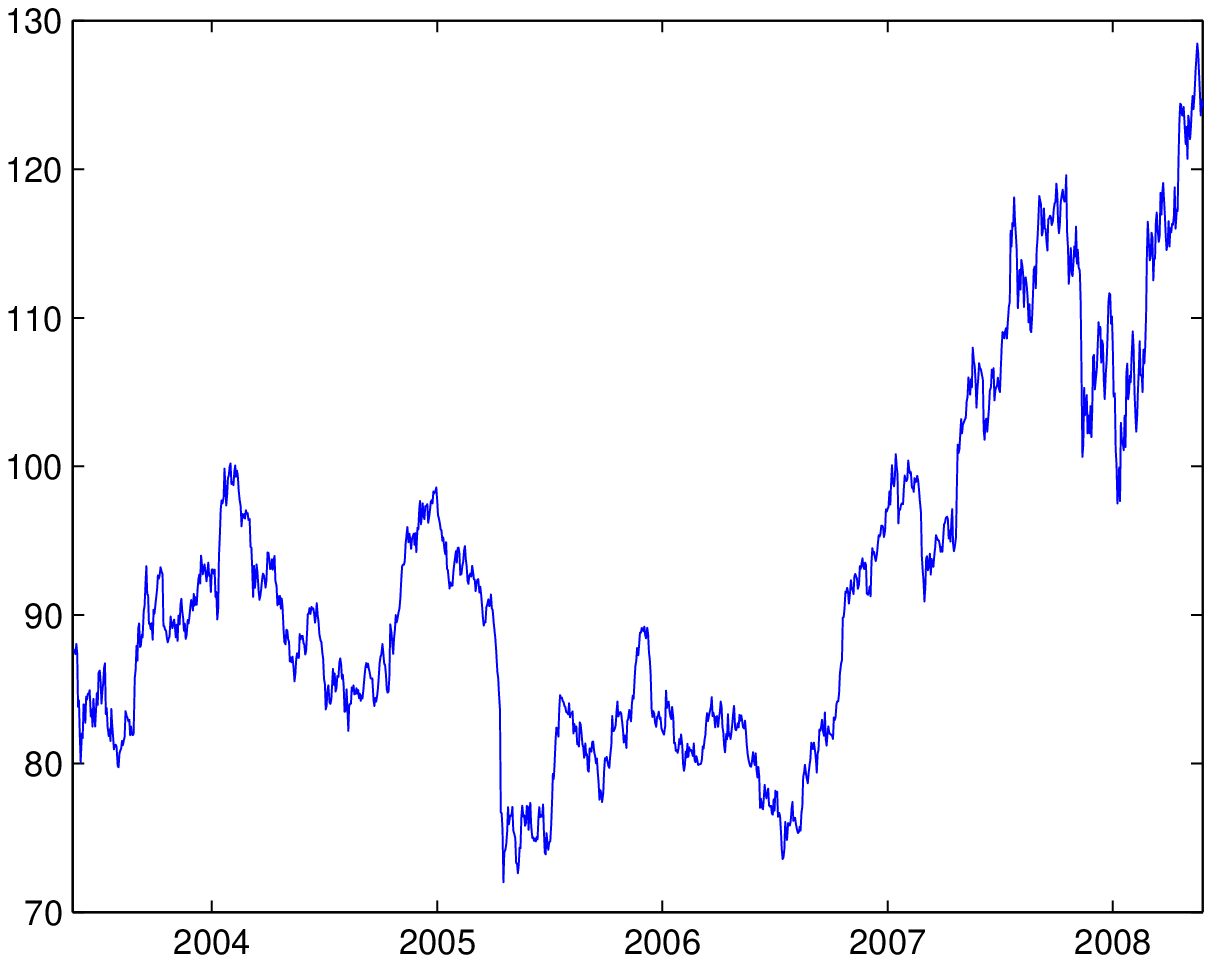} & \includegraphics[width=7.5cm,height=7cm]{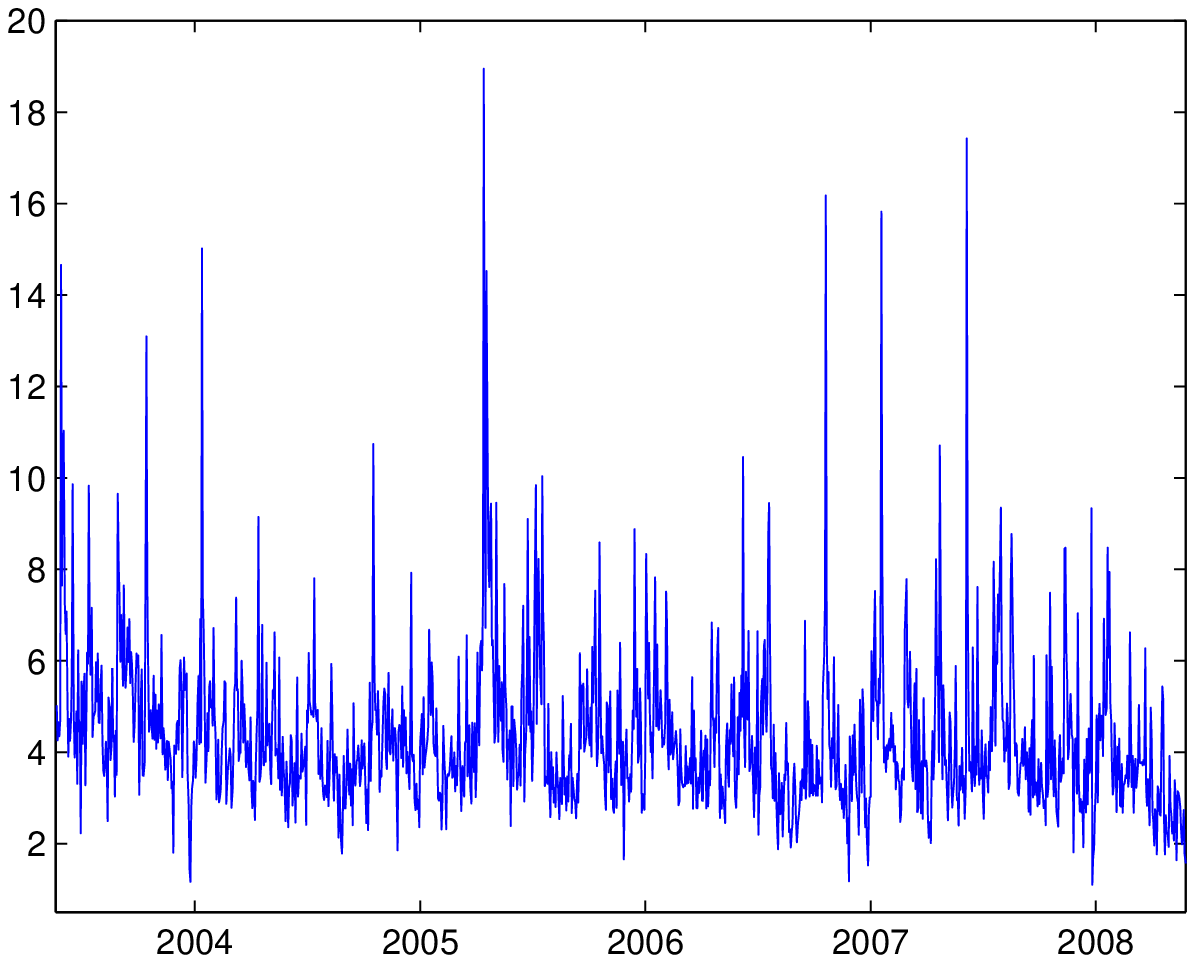}\\
\end{array}
$$
\caption{\emph{Left:} Closing prices for the IBM stock from March
$23$, $2003$ to March $23,$ $2008.$ \emph{Right:} Trading volumes
for the IBM stock during the same time period.\label{data_IBM}}
\end{figure}

\begin{figure}[h]
$$
\begin{array}{cc}
\includegraphics[width=7cm,height=4.5cm]{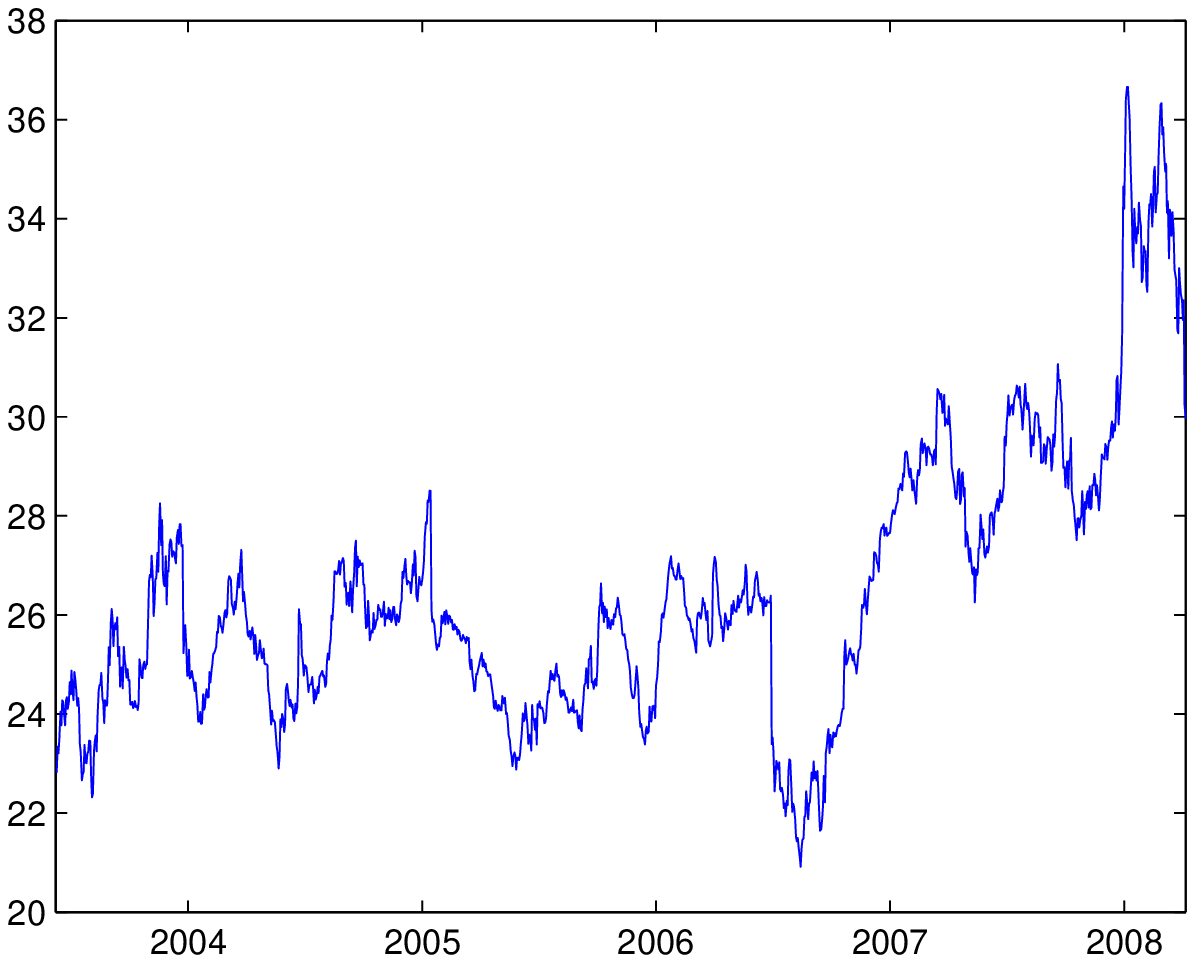} &
\includegraphics[width=7cm,height=4.5cm]{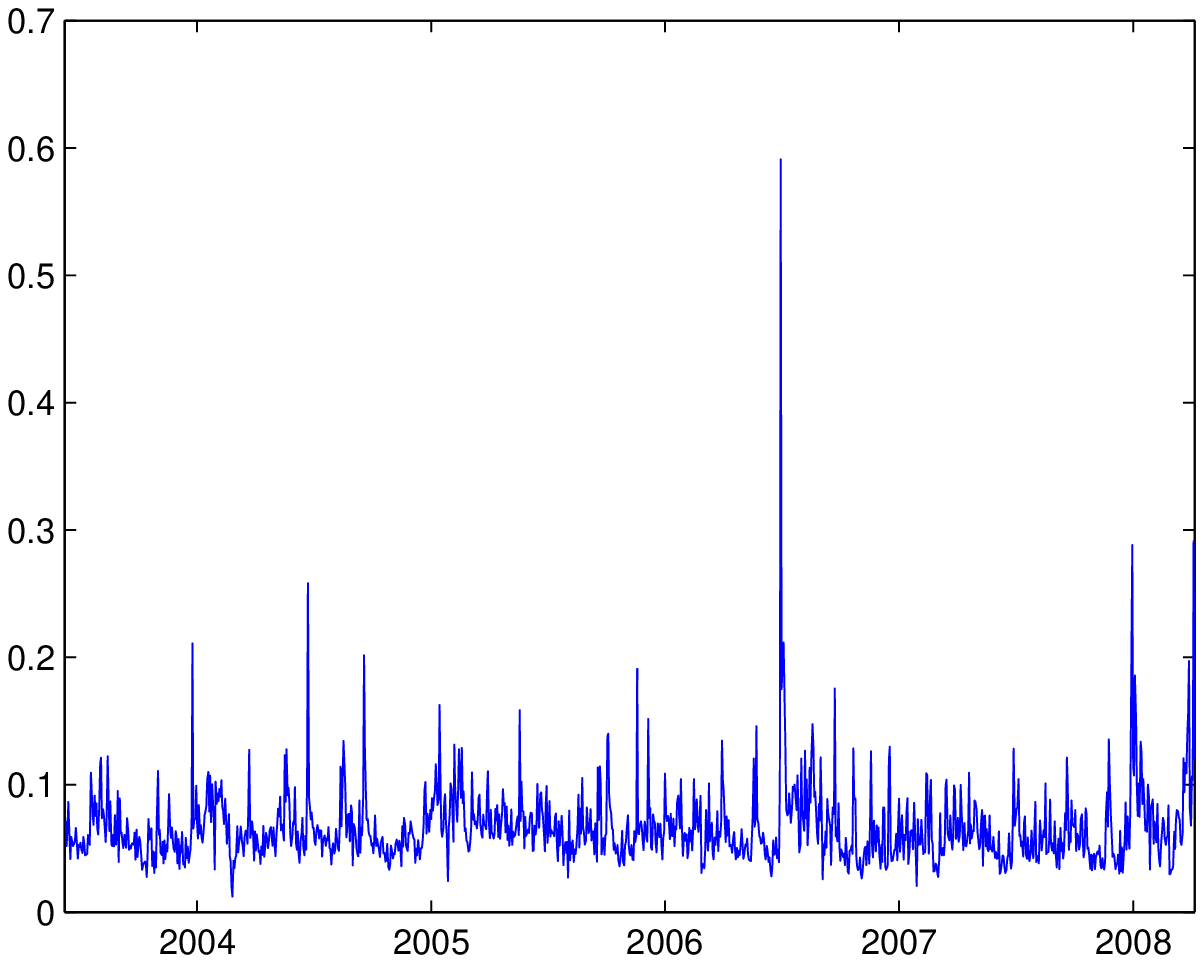}\\
\end{array}
$$
\caption{\emph{Left:} Closing prices for the MSFT stock from April
$11$, $2003$ to February $4,$ $2008.$ \emph{Right:} Trading
volumes for the MSFT stock during the same time
period.\label{data_MSFT}}
\end{figure}

The BNS model will be fitted to daily log returns of the
International Business Machines Corporation (IBM) stock and the
Microsoft Corporation (MSFT) stock. The IBM stock is from the New
York Stock Exchange (NYSE), whereas MSFT belongs to Nasdaq. 
The data spans over roughly 5 years starting in March $23$, $2003$
to March $23$, $2008$ for IBM and starting in April $11$, $2003$
to February $4$, $2008$ for MSFT. There were $1259$ and $1212$
observations for IBM and MSFT respectively of daily closing prices
and trading volumes. The resulting time series are shown in
Figure~\ref{data_IBM} and Figure~\ref{data_MSFT}. Data on trading
volumes are expressed in millions. Sample measures of skewness and
kurtosis\footnote{Skewness is measured as $\mu_3/\sqrt{\mu_2^3},$
and kurtosis as $\mu_4/\mu_2^2,$ where $\mu_i$ is the \emph{i}th
central moment.} of the returns are $-0.35$ and $7.42$
respectively for IBM and $-0.68$ and $14.2$ for MSFT. Throughout,
we consider the $\Gamma$-OU model from section~\ref{Sec-GaOU}.
\subsection{Parameter estimates and interpretation}
Table~\ref{Table6} presents the estimated parameter values for the
IBM and MSFT stocks. In the IBM case, for example, the parameters
imply that there are on average $4.4$ jumps per day each with mean
and standard deviation $0.704$. Typical volatility is $0.18$ with
standard deviation $0.11.$ The proportionality of trading volume
and the instantaneous variance is given by $\sigma^2=0.0076.$ The
leverage $\rho$ is very small.

\subsection{Returns distribution}
\begin{figure}
\begin{center}
\includegraphics[width=14cm,height=4.5cm]{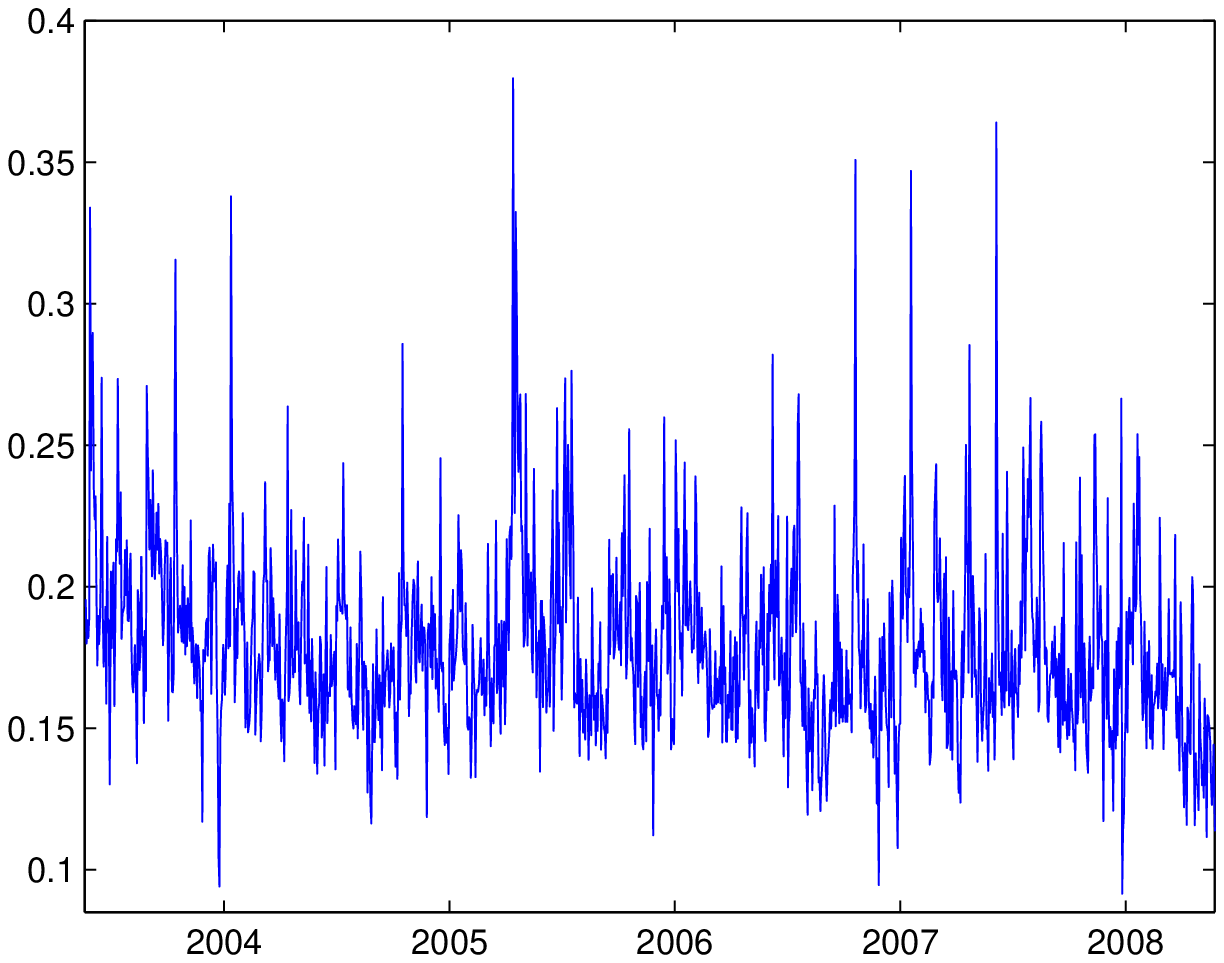}\\
\includegraphics[width=14cm,height=4.5cm]{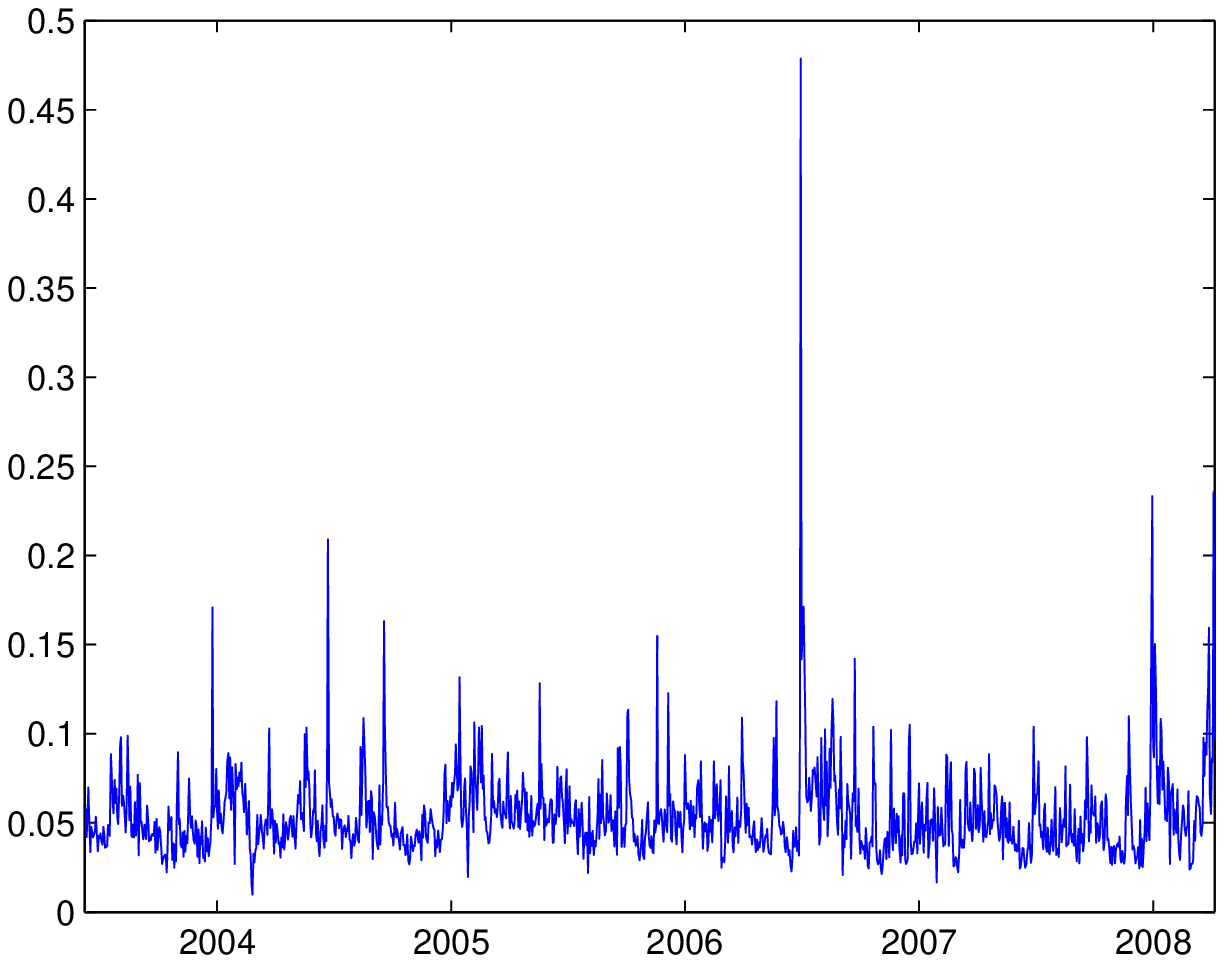}\\
\end{center}
\caption{The estimated volatility process
$\hat{\sigma}\times$$\sqrt{\tau}$ for IBM (\emph{top}) and MSFT
(\emph{bottom}).\label{Es_vlt}}
\end{figure}

\begin{figure}
$$
\begin{array}{cc}
\includegraphics[width=7.3cm,height=6.4cm]{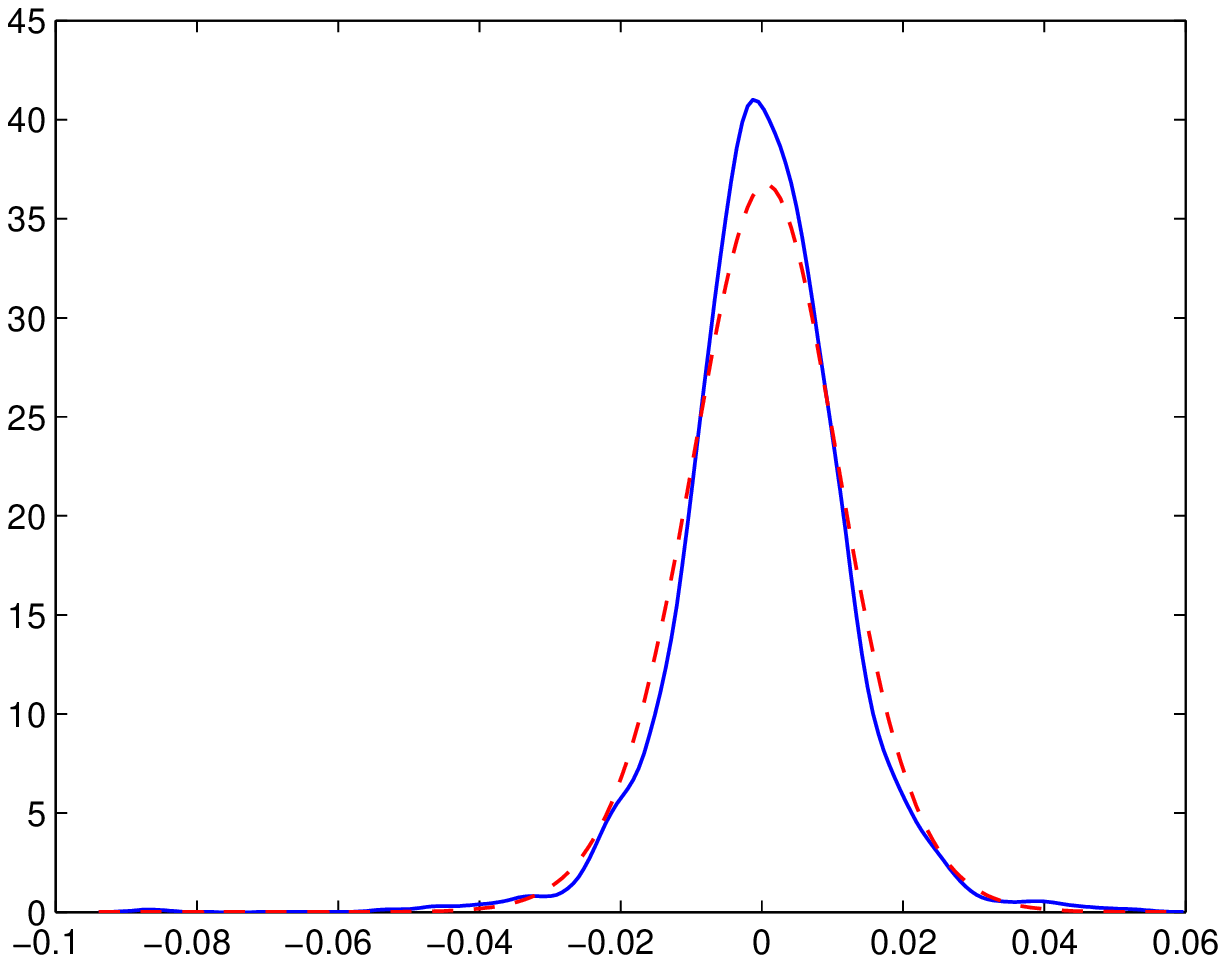} & \includegraphics[width=7.3cm,height=6.4cm]{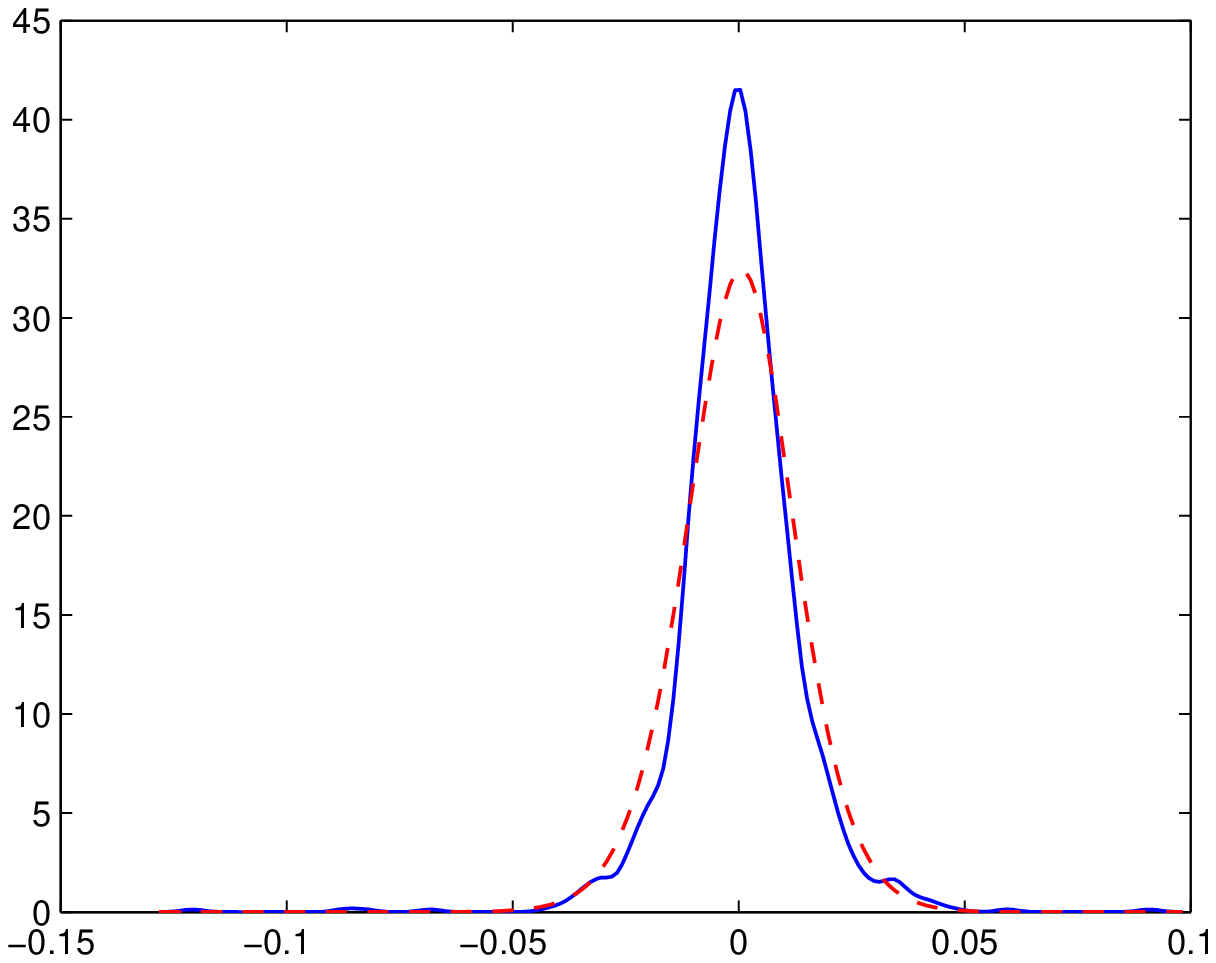}
\end{array}
$$
\caption{Theoretical densities (dashed line) and kernel estimates
of the empirical ones (solid line) of log returns for IBM
(\emph{left}) and MSFT (\emph{right}).\label{densityr}}
\end{figure}

The estimated parameters in the IBM case, for example, imply that
the mean of daily log-returns including or not a leverage effect
in the model equal $-0.027\%,$ and $0.146\%$ respectively. If the
trading volume process jumps by a typical size, the returns jump
by $0.0004.$ Using the estimated parameters, the volatility
processes for the IBM and MSFT stocks are shown in
Figure~\ref{Es_vlt}. In Table~\ref{Table7} some results on the
marginal moments of daily log returns and the instantaneous
variance process $V(t)$ using the estimated model parameters are
presented. Furthermore, the theoretical density and log density of
log returns and the estimated ones are shown in
Figure~\ref{densityr} and Figure~\ref{logdensityr}. A systematic
method how to calculate the theoretical density of log returns is
given in Appendix~\ref{App}.

\subsection{The autocorrelation function}
\begin{figure}[h]
$$
\begin{array}{cc}
\includegraphics[width=7.3cm,height=6.4cm]{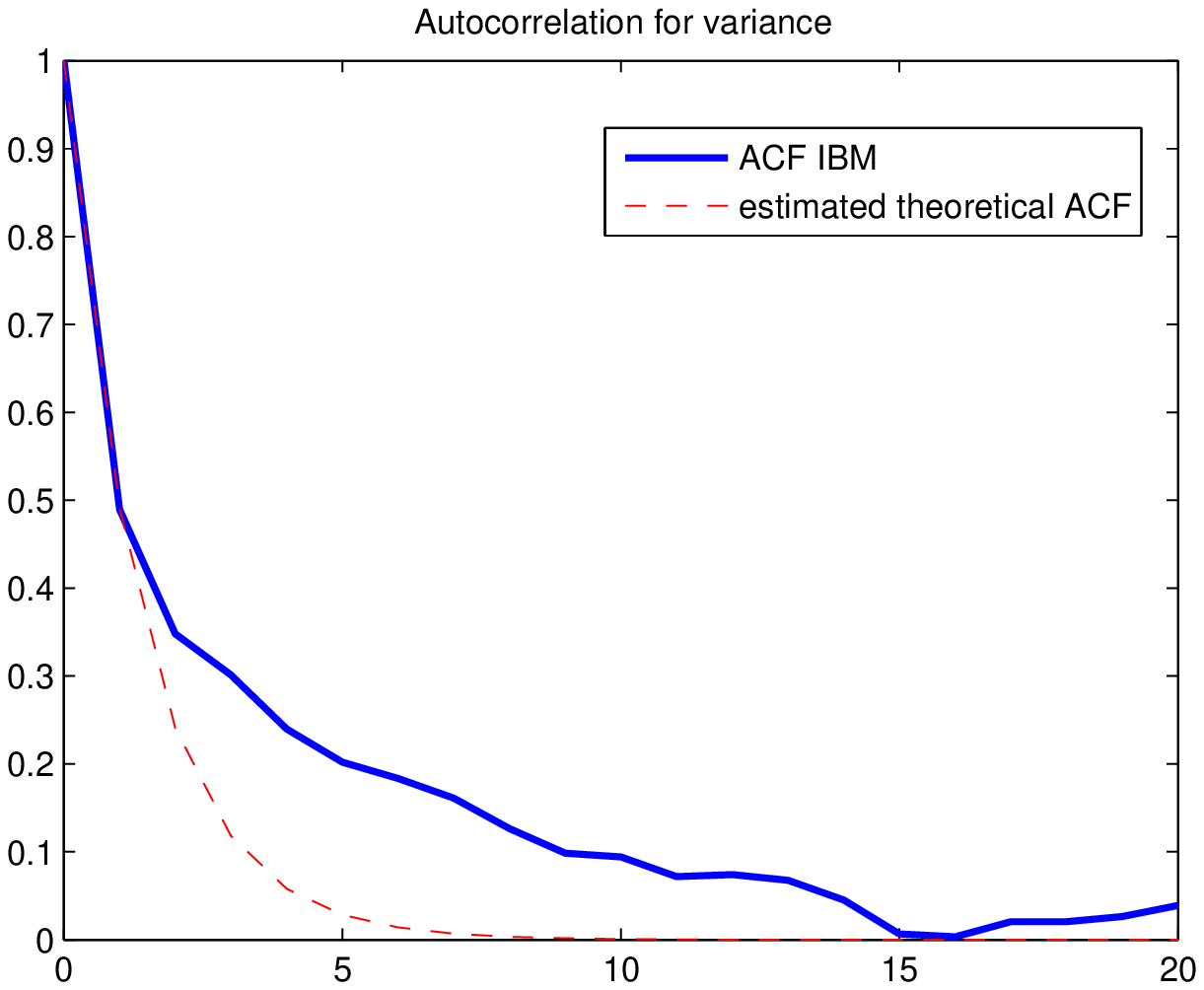} & \includegraphics[width=7.3cm,height=6.4cm]{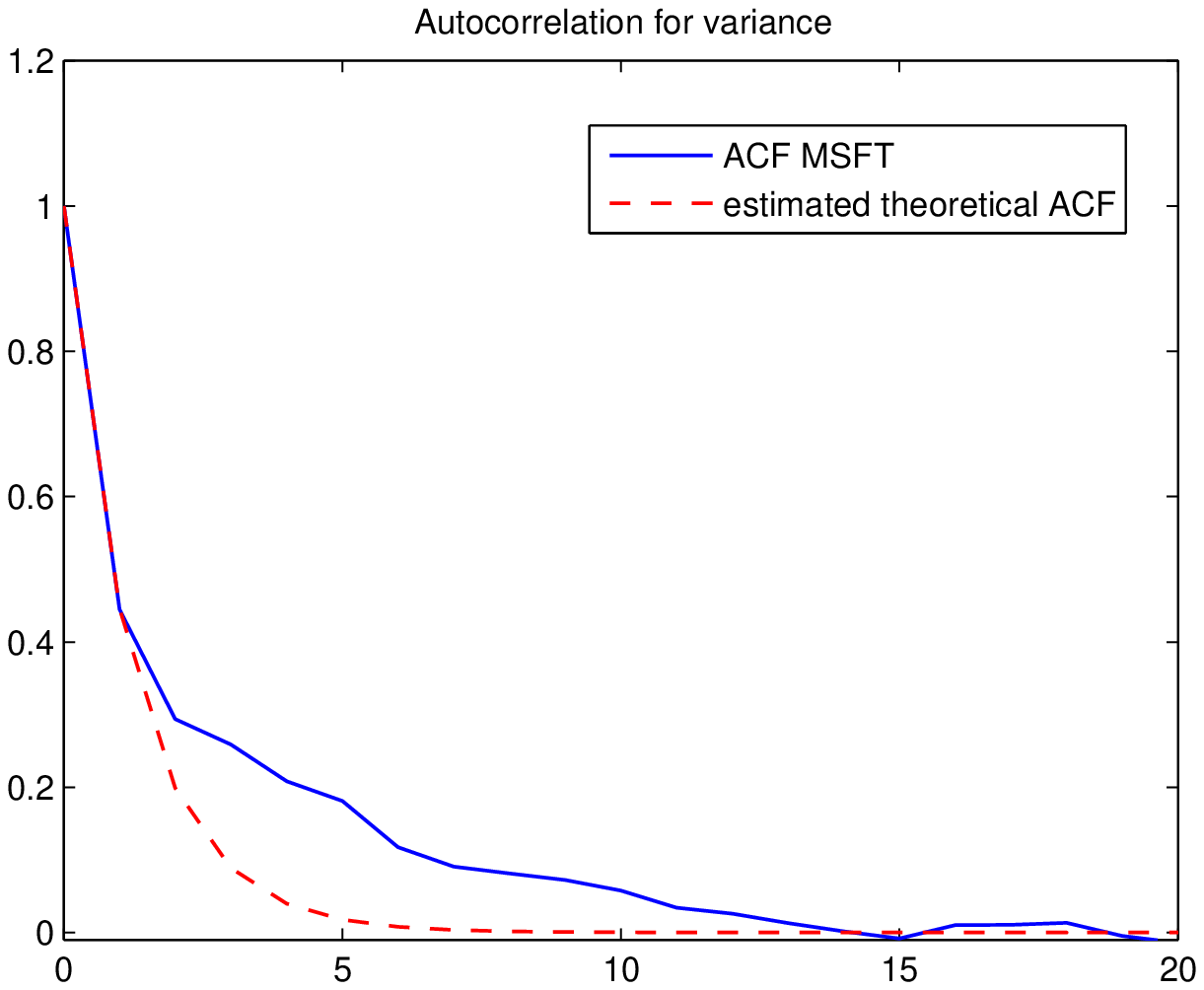}\\
\end{array}
$$
\caption{The autocorrelation function for the variance process and
the estimated theoretical autocorrelation for IBM (\emph{left})
and for MSFT (\emph{right}).\label{ACF}}
\end{figure}

The theoretical autocorrelation function for the variance process
and the estimated autocorrelation for both the stocks are shown in
Figure \ref{ACF} which is not very satisfactory. We will address
this issue in the concluding remarks below.
\begin{table}
\begin{center}
\begin{tabular}{|c||c||c|}
\hline Parameter & Value IBM & St.dev.\\
\hline $\hat{\nu}$ & 6.17 & 0.339\\
\hline $\hat{\alpha}$ & 1.42 & 0.079\\
\hline $\hat{\lambda}$ & 177.95 & 12.509\\
\hline $\hat{\mu}$ & 0.435 & 0.254\\
\hline $\hat{\beta}$ & -0.015 & 0.072\\
\hline $\hat{\sigma}$ & 0.087 & 0.002\\
\hline $\hat{\rho}$ & -0.00056 & 0.0002\\
\hline
\end{tabular}\qquad
\begin{tabular}{|c||c||c|}
\hline Parameter & Value MSFT & St.dev.\\
\hline $\hat{\nu}$ & 4.496 & 0.247\\
\hline $\hat{\alpha}$ & 67.895 & 3.773\\
\hline $\hat{\lambda}$ & 201.99 & 14.420\\
\hline $\hat{\mu}$ & 0.4162 & 0.265\\
\hline $\hat{\beta}$ & -0.464 & 5.059\\
\hline $\hat{\sigma}$ & 0.81 & 0.018\\
\hline $\hat{\rho}$ & -0.025 & 0.013\\
\hline
\end{tabular}
\caption{Estimated parameter values.\label{Table6}}
\end{center}
\end{table}
\begin{table}
\begin{center}
\begin{tabular}{|c||c|}
\hline  Unconditional moments IBM & Value \\
\hline $E[X]$ & $-0.027\%$ \\
\hline $St.dev[X]$ & $1.15\% $ \\
\hline $E[V]$ & $3.3\%$\\
\hline $St.dev[V]$ & $1.32\%$\\
\hline
\end{tabular}
\qquad
\begin{tabular}{|c||c|}
\hline  Unconditional moments MSFT & Value \\
\hline $E[X]$ & $0.02\%$ \\
\hline $St.dev[X]$ & $1.32\% $ \\
\hline $E[V]$ & $4.34\%$\\
\hline $St.dev[V]$ & $2.05\%$\\
\hline
\end{tabular}
\caption{Unconditional moments calculated from the estimated
parameters.\label{Table7}}
\end{center}
\end{table}

\subsection{The model fit}
\begin{figure}
$$
\begin{array}{cc}
\includegraphics[width=7.3cm,height=6.4cm]{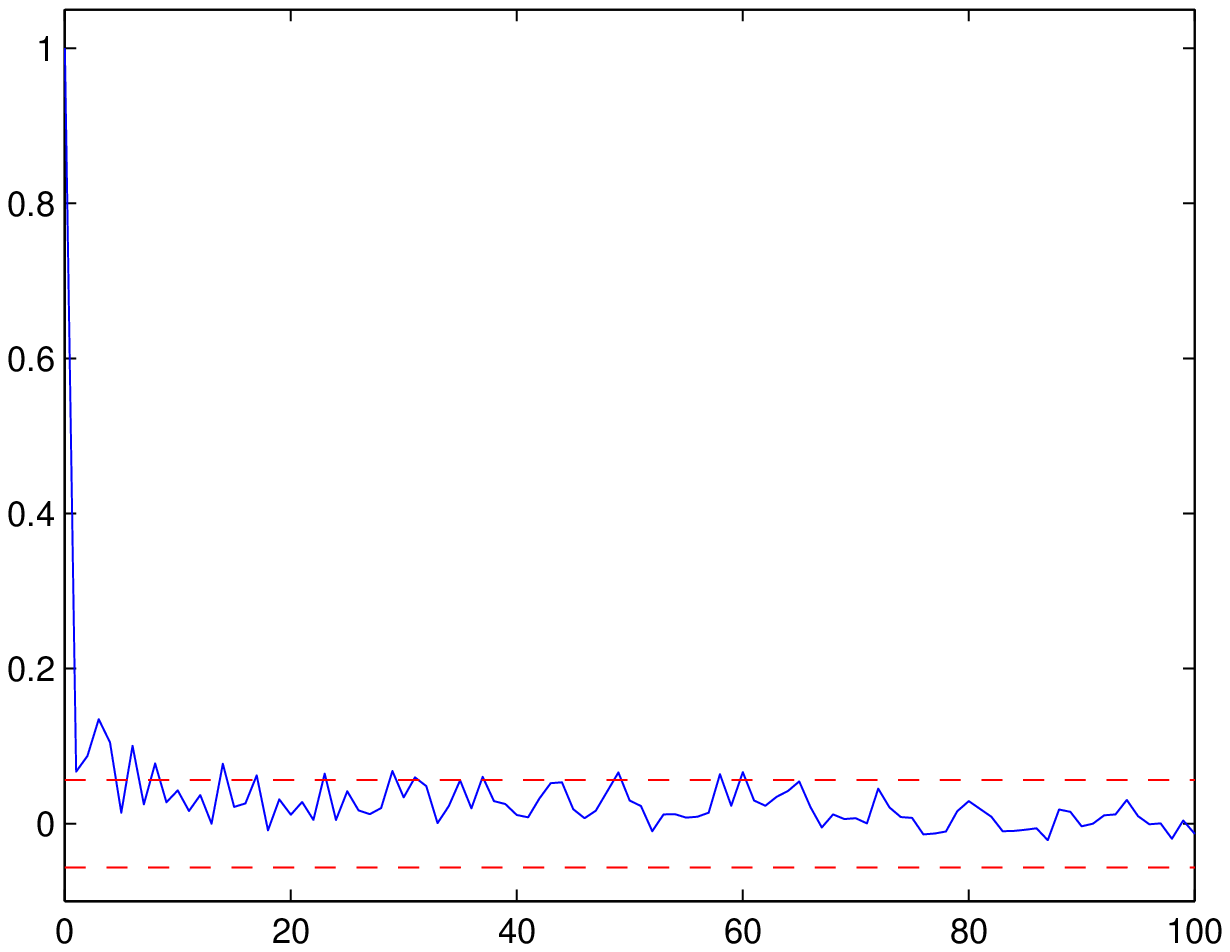} & \includegraphics[width=7.3cm,height=6.4cm]{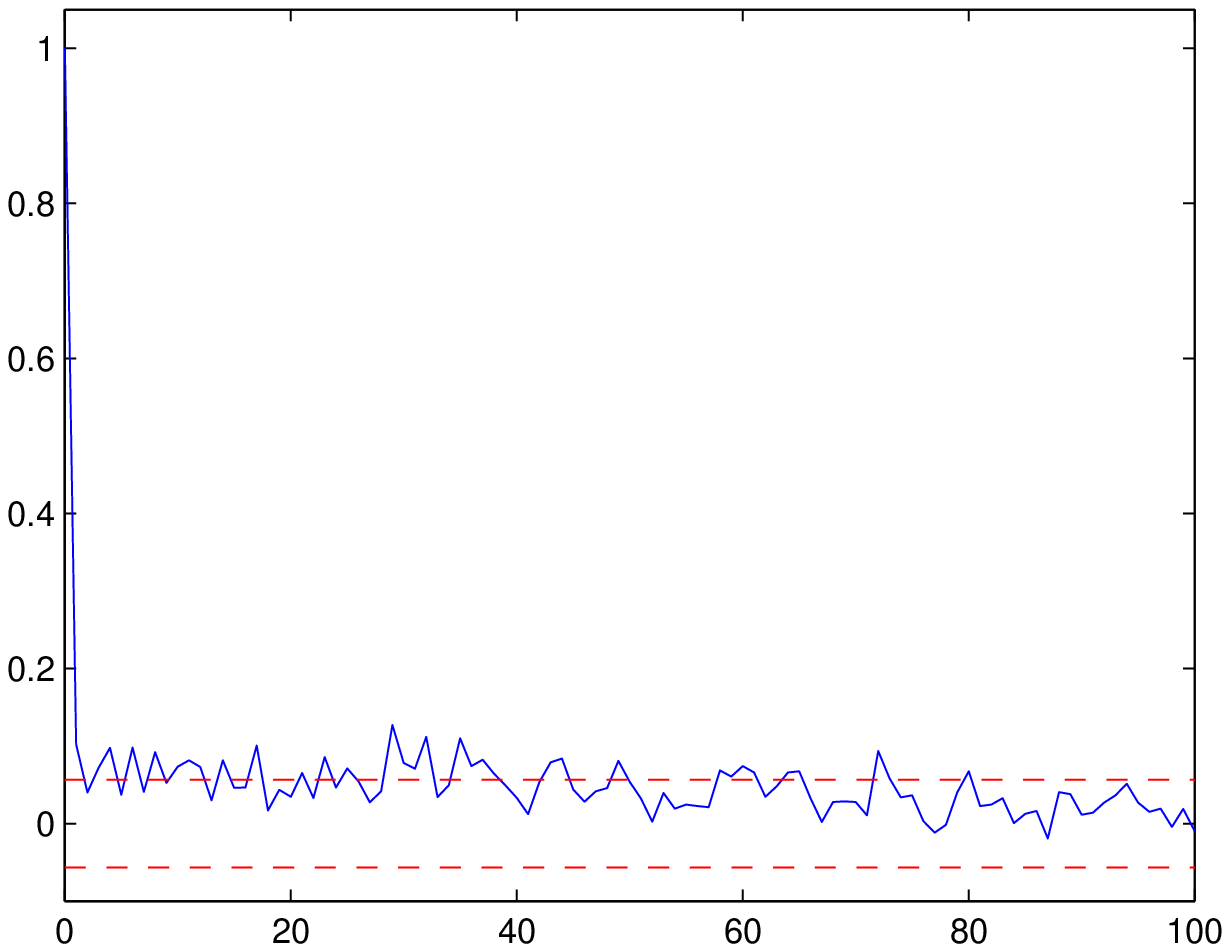}
\end{array}
$$
\caption{\emph{Left:}~The empirical autocorrelation function for
the squared log returns for IBM during the period March 23, 2003
to March 23, 2008. \emph{Right:}~The empirical autocorrelation
function for the squared residuals for IBM during the same period.
The straight lines are the asymptotic $95\%$ confidence bands $\pm
1.96\sqrt{n},$ where $n$ is the number of
observations.\label{Sq_residuals}}
\end{figure}

\begin{figure}
$$
\begin{array}{cc}
\includegraphics[width=7.3cm,height=6.4cm]{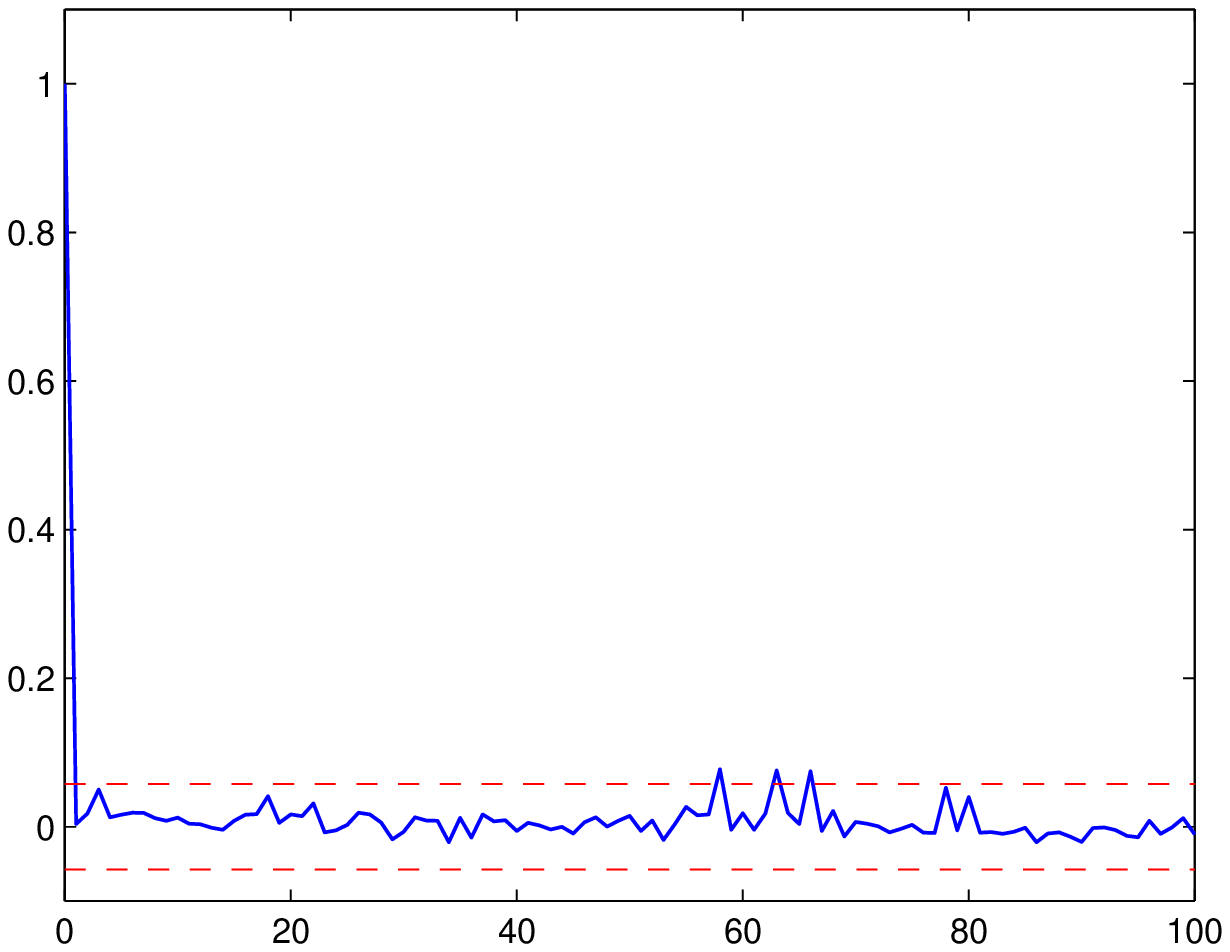} & \includegraphics[width=7.3cm,height=6.4cm]{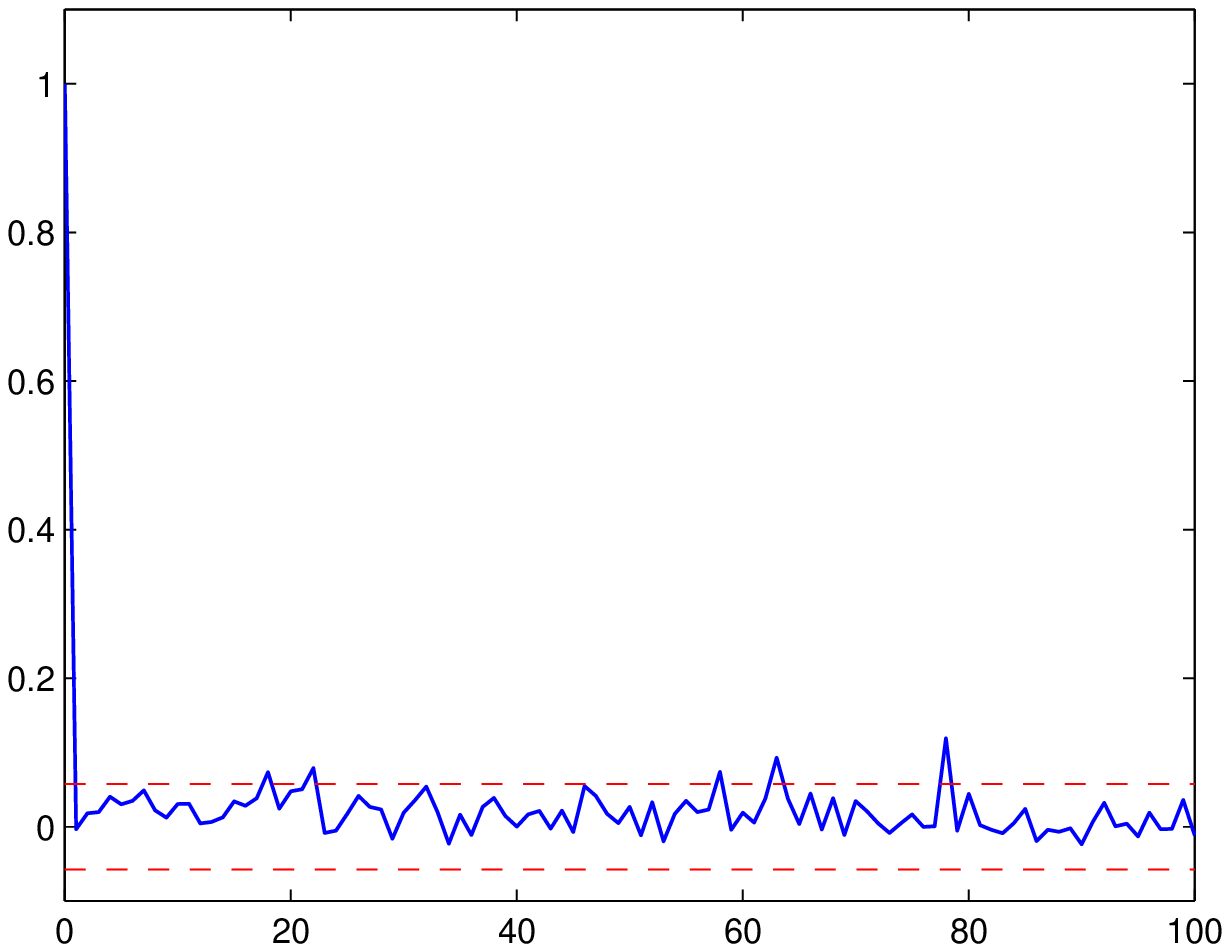}
\end{array}
$$
\caption{\emph{Left:}~The empirical autocorrelation function for
the squared log returns for MSFT during the period April 11, 2003
to February 4, 2008. \emph{Right:}~The empirical autocorrelation
function for the squared residuals for MSFT during the same
period. The straight lines are the asymptotic $95\%$ confidence
bands $\pm 1.96\sqrt{n},$ where $n$ is the number of
observations.\label{Sq_residuals_MSFT}}
\end{figure}

\begin{figure}
$$
\begin{array}{cc}
\includegraphics[width=7.3cm,height=6.4cm]{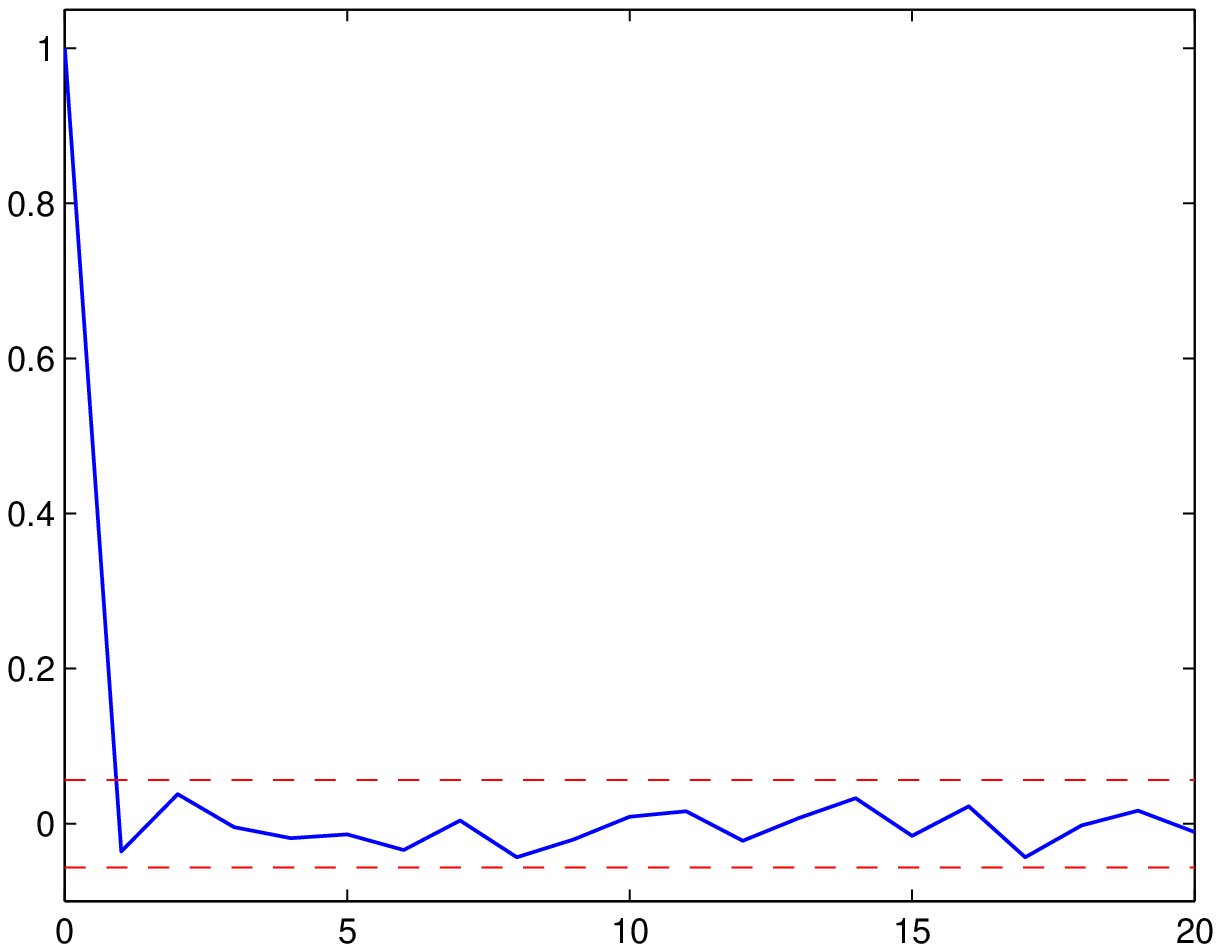} & \includegraphics[width=7.3cm,height=6.4cm]{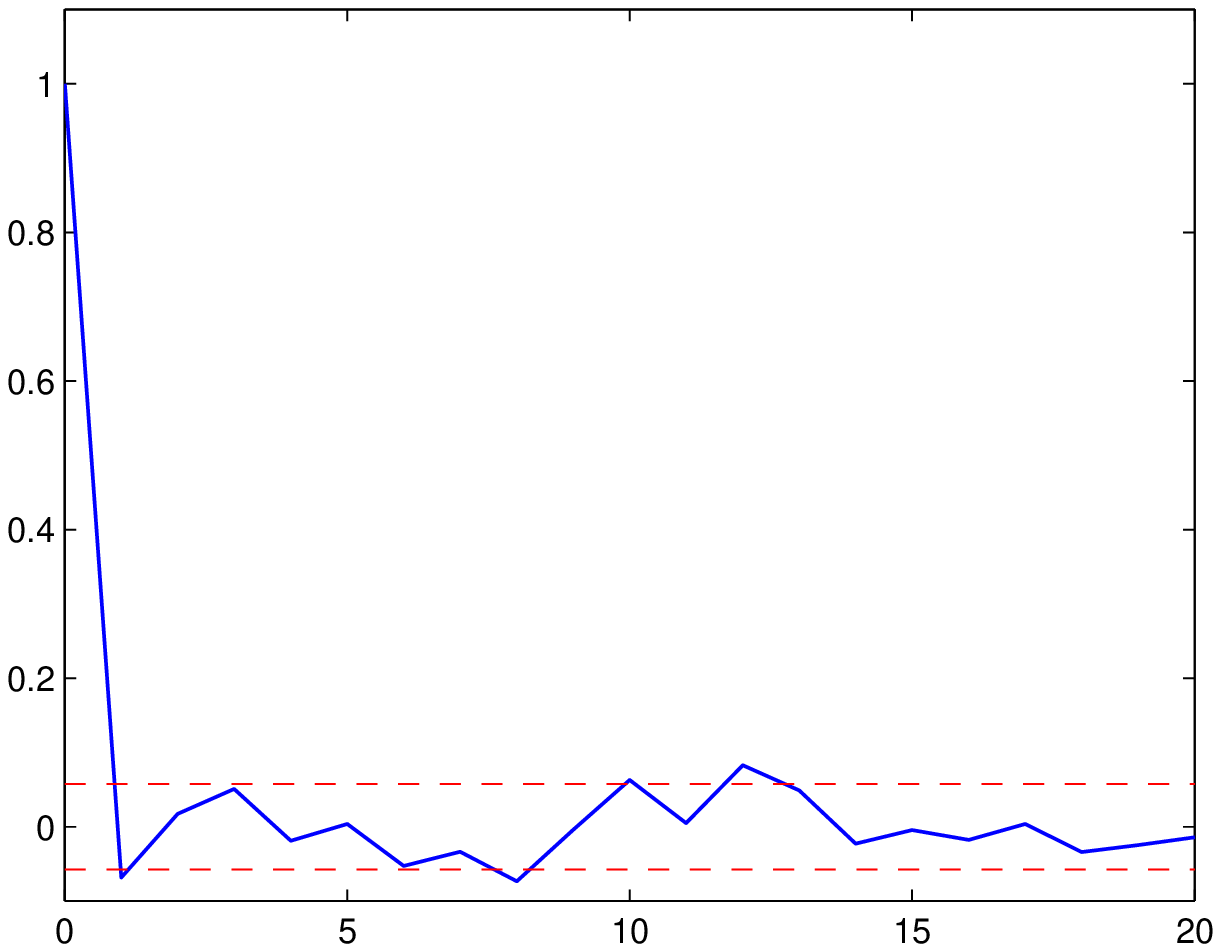}
\end{array}
$$
\caption{The autocorrelation function for the \emph{residuals} for
IBM (\emph{left}) during the period March 23, 2003 to March 23,
2008 and MSFT (\emph{right}) during the period April, $11,$ $2003$
to February $4,$ $2008$ .\label{ACF_norm_returns}}
\end{figure}

To investigate the model fit, we performed a Ljung-Box test for
squared residuals of the data set. The test statistic used $35$
lags of the corresponding empirical autocorrelation function. The
 null hypothesis was not rejected for MSFT at the $0.05$ level.
 For the MSFT squared residuals the \emph{p}-value was $0.052$
 The test statistic for the IBM squared residuals was equal to
$451.61,$ which led to a rejection of the null hypothesis, since
the test had a critical value of $113.15$ at the $0.05$ level.
This result is also obvious from Figure \ref{Sq_residuals} where
the empirical autocorrelation function of the squared residuals is
plotted, showing significant correlations of the IBM residuals.
The empirical autocorrelation functions of the squared residuals
for MSFT is shown in Figure~\ref{Sq_residuals_MSFT}. Furthermore,
the autocorrelation functions of $\hat{\varepsilon}(\cdot)$ for
both the stocks are shown
in Figure~\ref{ACF_norm_returns}.

\begin{figure}
$$
\begin{array}{cc}
\includegraphics[width=7.3cm,height=6.4cm]{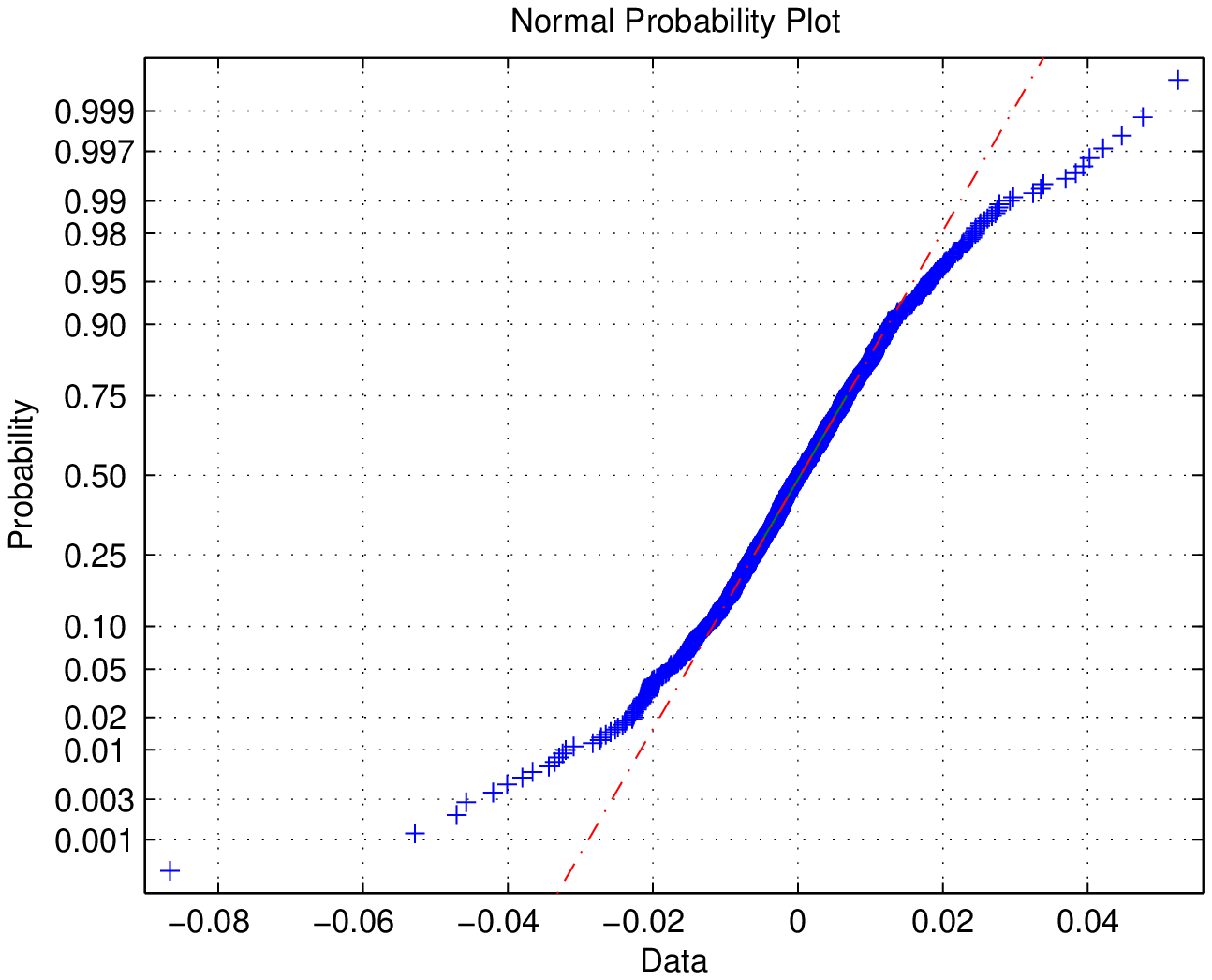} & \includegraphics[width=7.3cm,height=6.4cm]{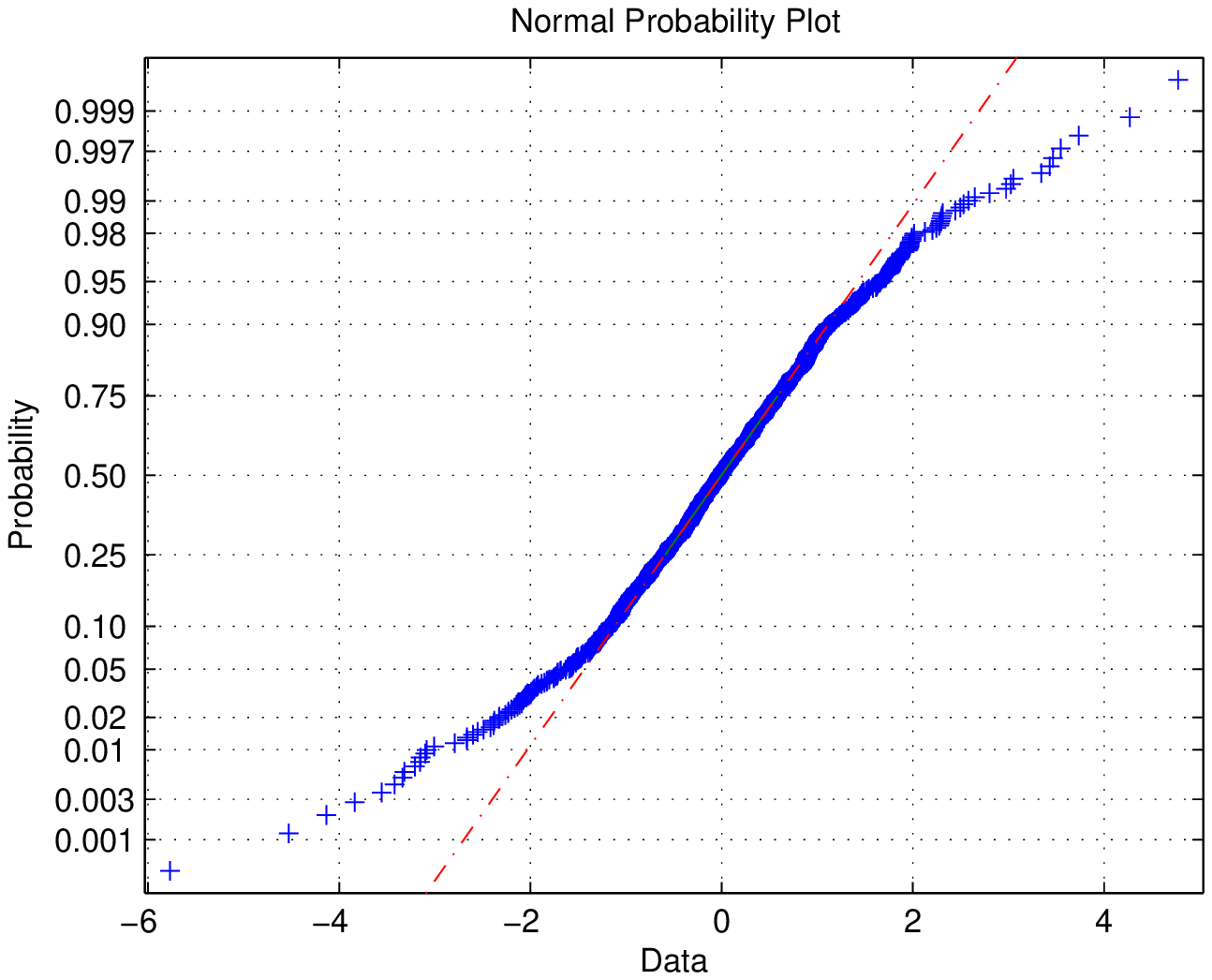}
\end{array}
$$
\caption{\emph{Left:}~The normal probability plot of log returns
for IBM during the period March 23, 2003 to March 23, 2008.
\emph{Right:}~The normal probability plot of \emph{residuals}
 for IBM during the same period.\label{npps}}
\end{figure}

\begin{figure}
$$
\begin{array}{cc}
\includegraphics[width=7.3cm,height=6.4cm]{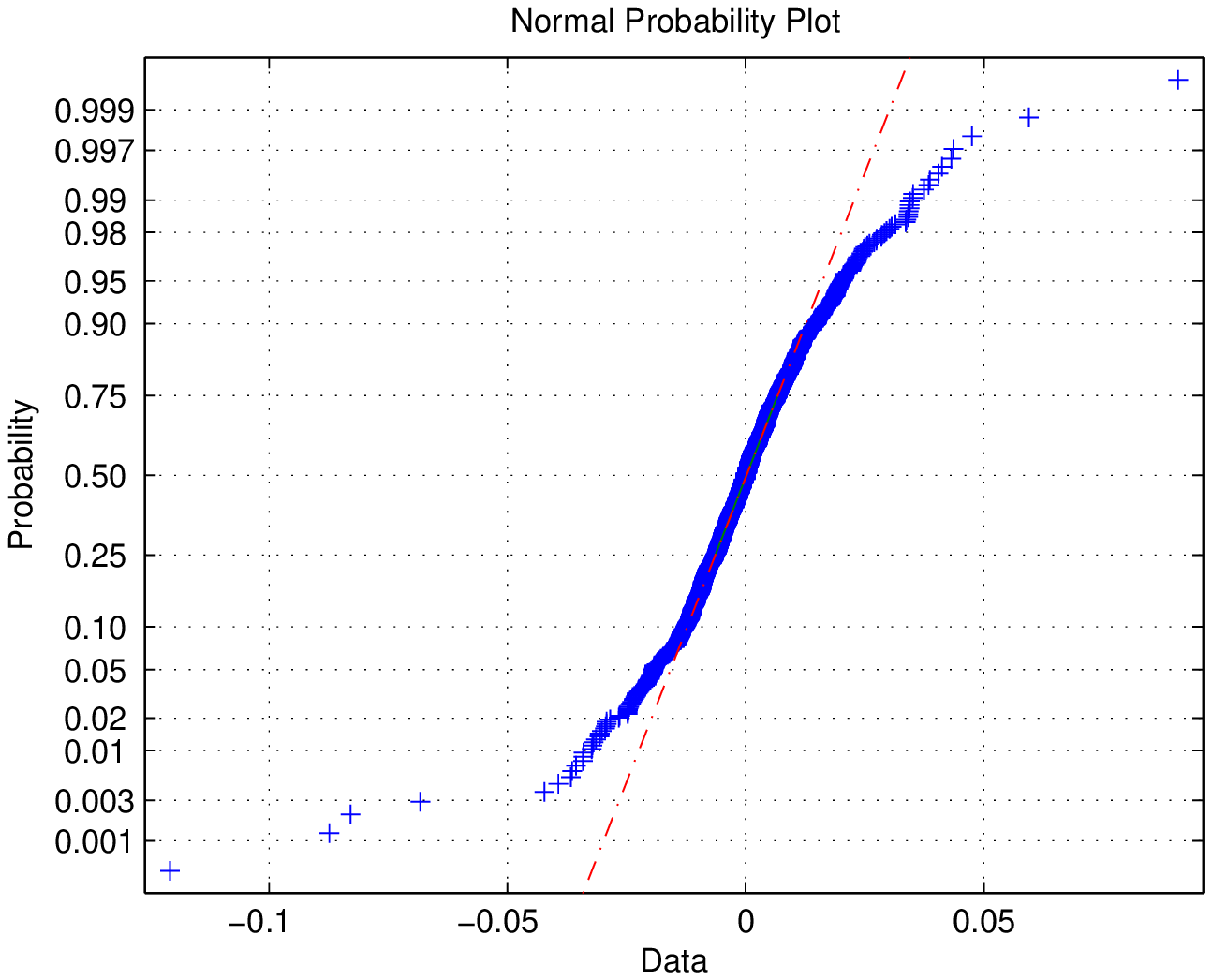} & \includegraphics[width=7.3cm,height=6.4cm]{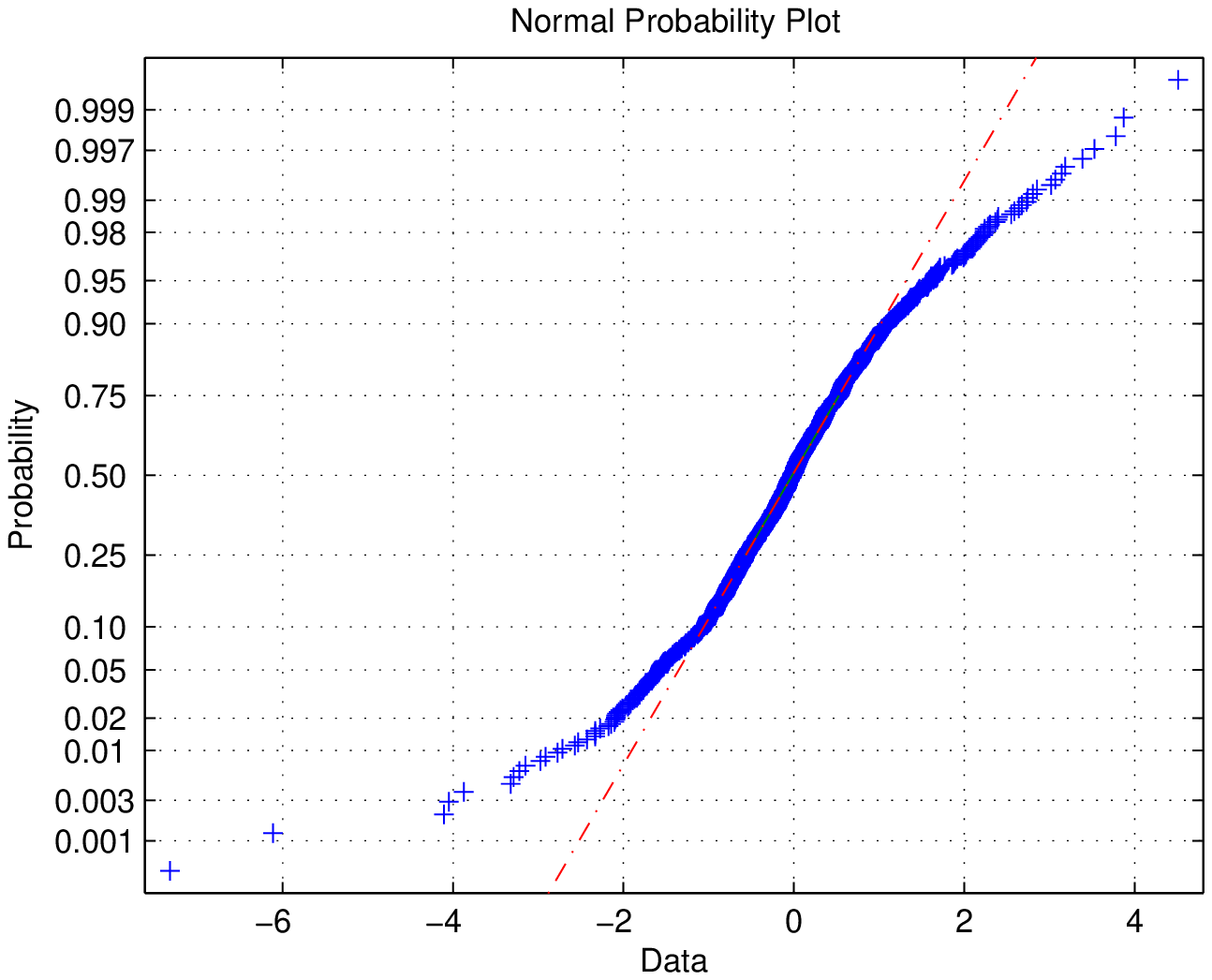}
\end{array}
$$
\caption{\emph{Left:}~The normal probability plot of log returns
for MSFT during the period April 11, 2003 to February 4, 2008.
\emph{Right:}~The normal probability plot of \emph{residuals}
 for MSFT during the same period.\label{npps_MSFT}}
\end{figure}

\begin{figure}
$$
\begin{array}{cc}
\includegraphics[width=7.3cm,height=6.4cm]{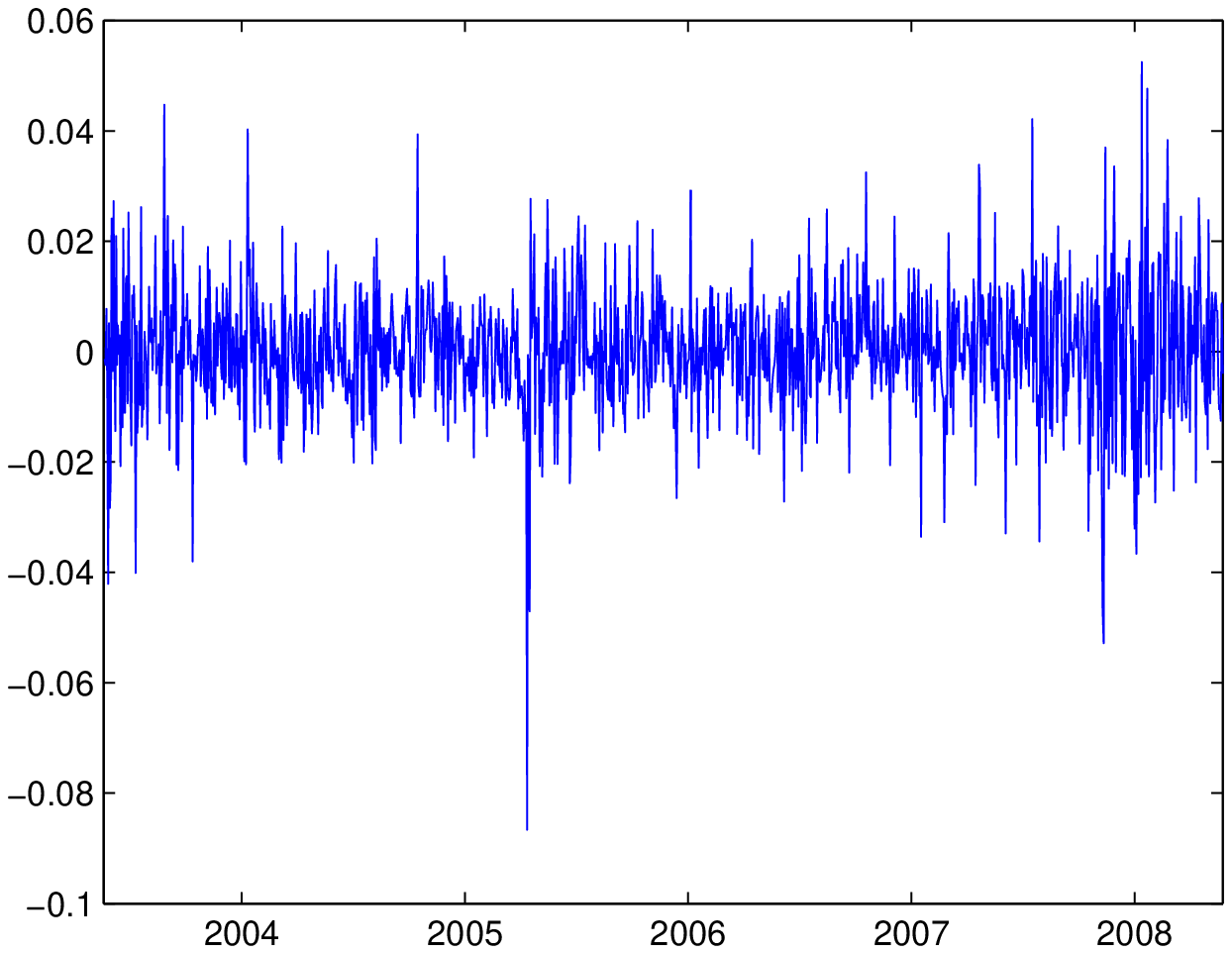} & \includegraphics[width=7.3cm,height=6.4cm]{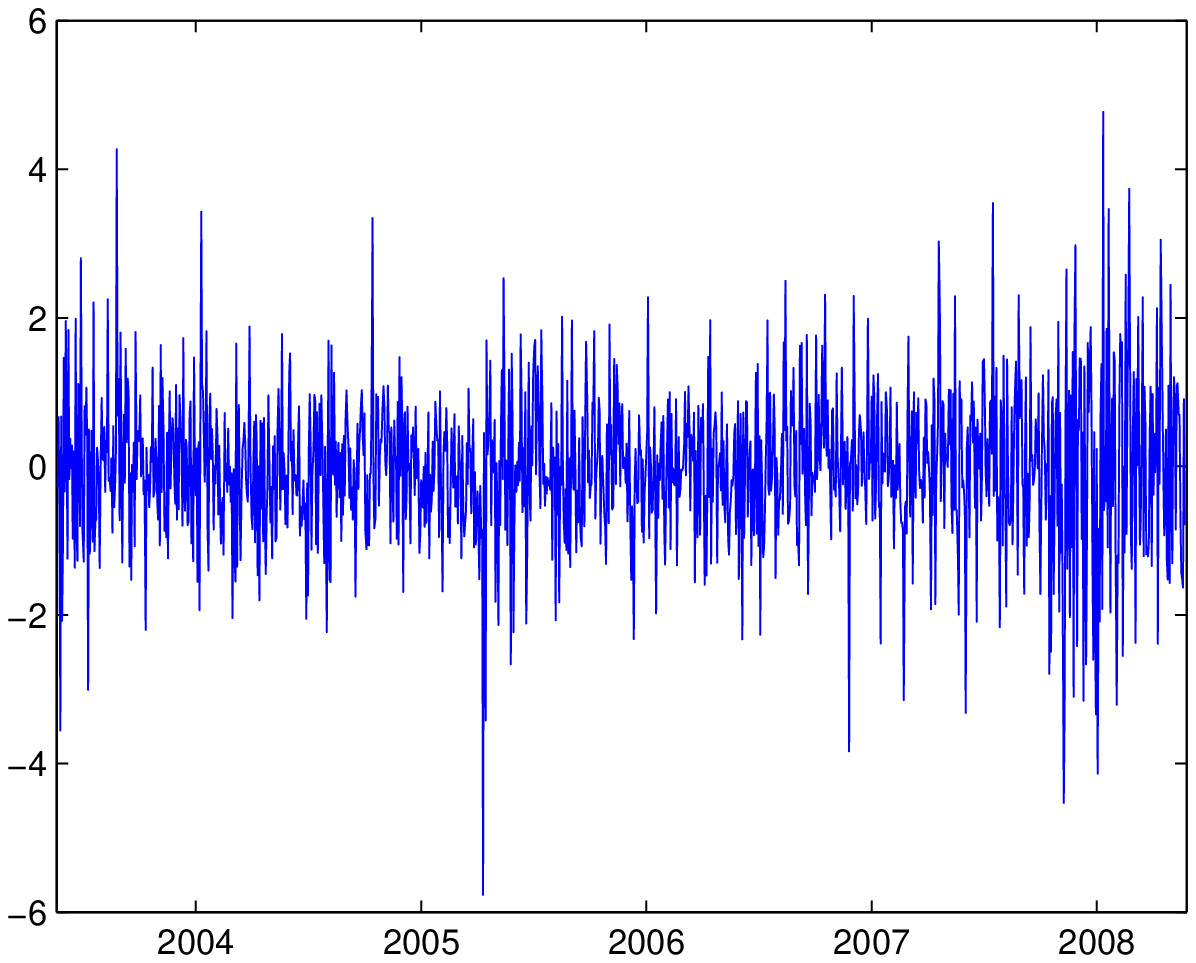}
\end{array}
$$
\caption{\emph{Left:}~Log returns for IBM during the period March
$23,$ $2003$ to March $23$, $2008$. \emph{Right:}~The
\emph{residuals} for IBM during the same time
period.\label{Returns_IBM}}
\end{figure}

\begin{figure}
$$
\begin{array}{cc}
\includegraphics[width=7.3cm,height=6.4cm]{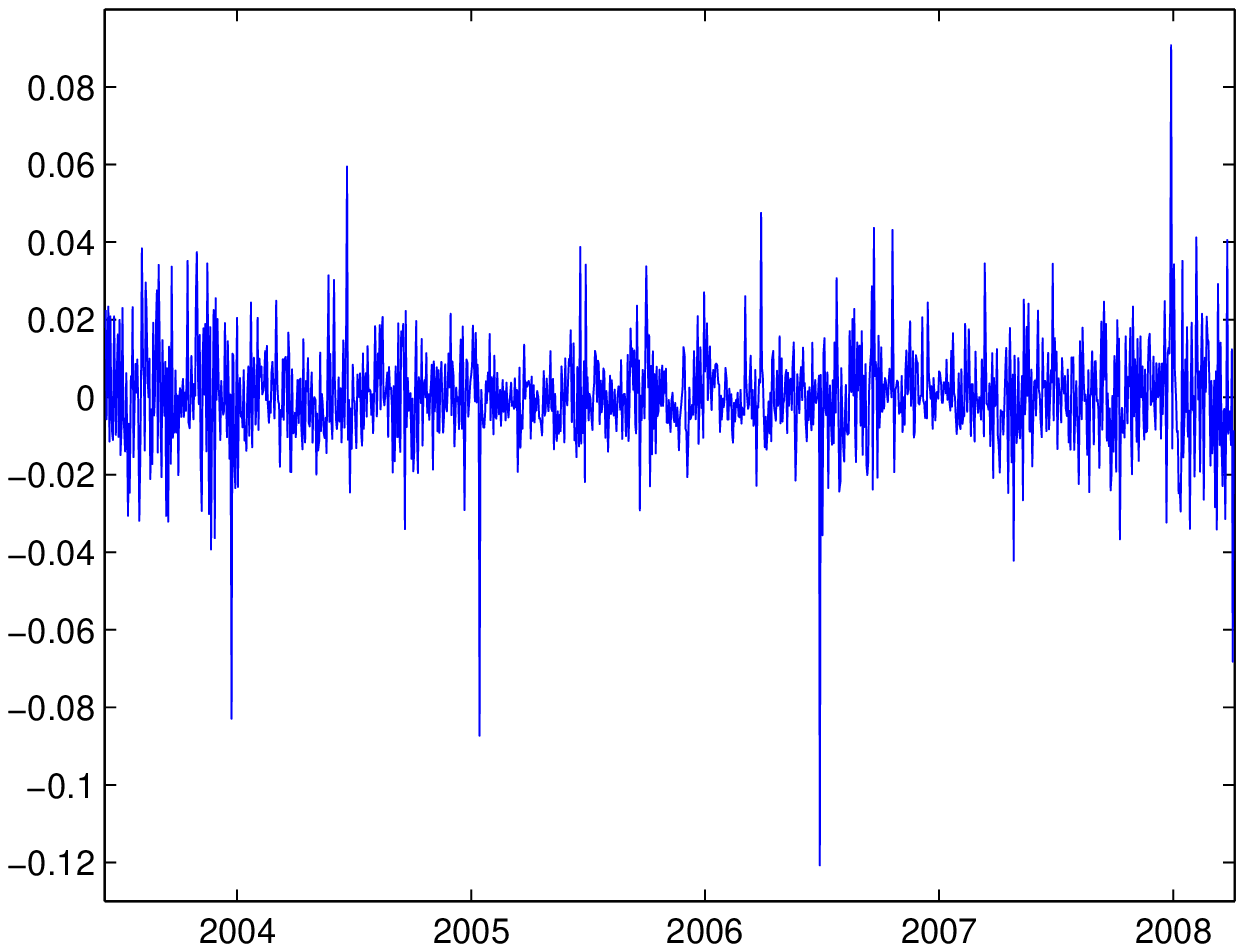} & \includegraphics[width=7.3cm,height=6.4cm]{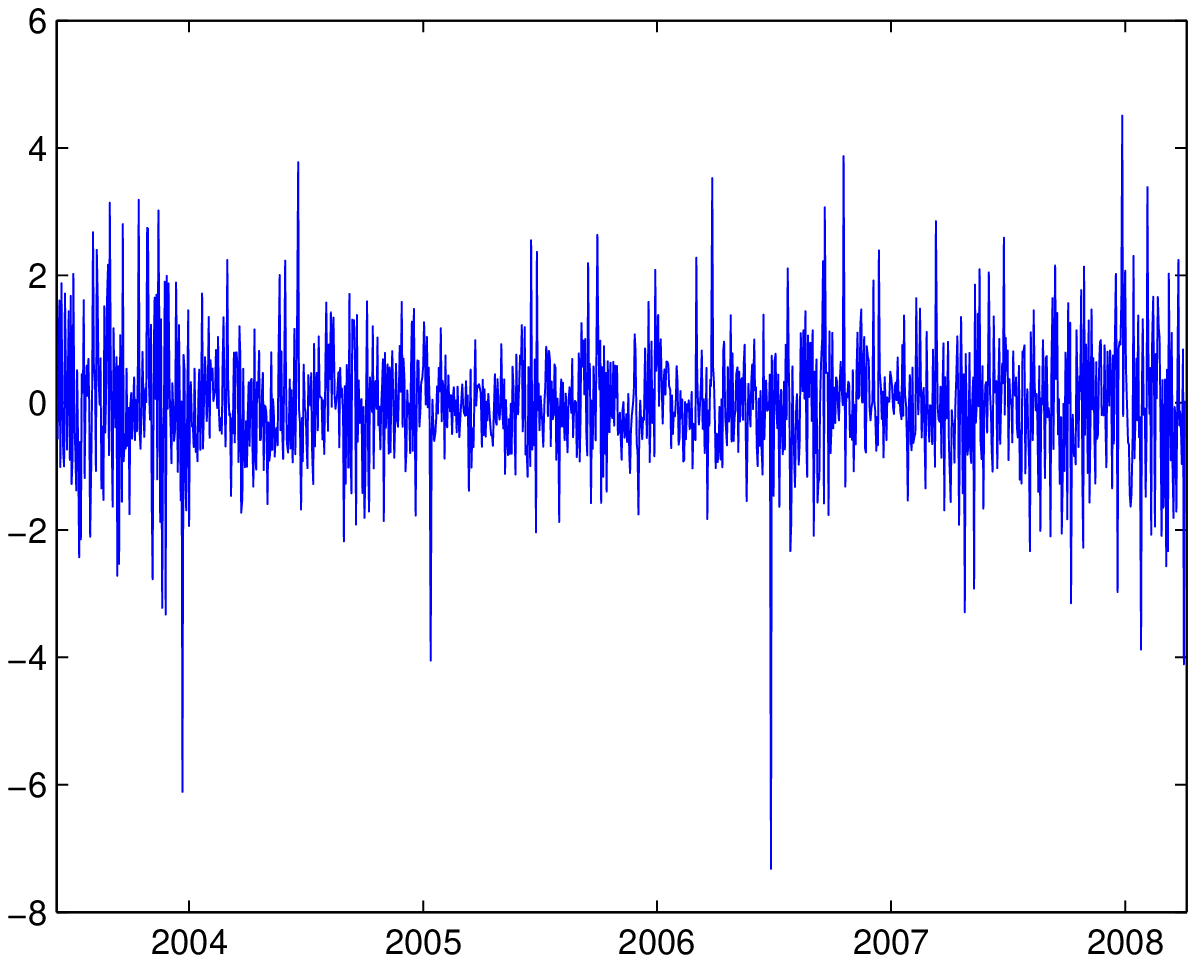}
\end{array}
$$
\caption{\emph{Left:}~Log returns for MSFT during the period April
$11,$ $2003$ to February $4$, $2008$. \emph{Right:}~The
\emph{residuals} for MSFT during the same time
period.\label{Returns_MSFT}}
\end{figure}

The estimated mean, standard deviation, skewness and kurtosis of
the residuals for both the stocks are summarized in Table
\ref{Table8}.
\begin{table}
\begin{center}
\begin{tabular}{|c|c|c|c|c|}
\hline & mean($\hat{\varepsilon}$) & std($\hat{\varepsilon}$)& skew($\hat{\varepsilon}$) & kurt($\hat{\varepsilon}$) \\
\hline IBM &  $-0.01568$ & $ 1.03142 $ & $-0.17053$ & $5.40659$  \\
\hline MSFT & $-0.01375$ & $ 1.00386 $ & $-0.35846$ & $7.86258$\\
\hline
\end{tabular}\caption{Mean, standard deviation, skewness and kurtosis of the IBM residuals.\label{Table8}}
\end{center}
\end{table}
The numbers show that the mean and variation of the residuals are
according to our model, but the residuals seem to have heavier
tails than
the normal distribution, see Figure~\ref{npps} and Figure~\ref{npps_MSFT}. 
The IBM residuals
 pass the Kolmogorov-Smirnov test of normality\footnote{In \cite{Lin2007}
 a similar approach is used, with number of trades as a measure of trading intensity. In that work, superposition
 of two
 OU-processes are analyzed for the modelling procedure, but without a leverage in the specification of returns. It is also pointed out that typically their
 approach gives normalized returns with heavier tails than the normal distribution for illiquid stocks and for
 special dates such as the trading day before a holiday.}, for example, with \emph{p}-value
 $0.0886,$ whereas the test statistic for the MSFT residuals was equal to $0.0622$, which
 lead to a rejection of the null  hypothesis, since the test had a
 critical value of $0.0389$ at the $0.05$ level.
 Log returns for the IBM and MSFT stocks and their
residuals are shown in Figure \ref{Returns_IBM} and Figure
\ref{Returns_MSFT}.
\begin{figure}
$$
\begin{array}{cc}
\includegraphics[width=7.3cm,height=6.4cm]{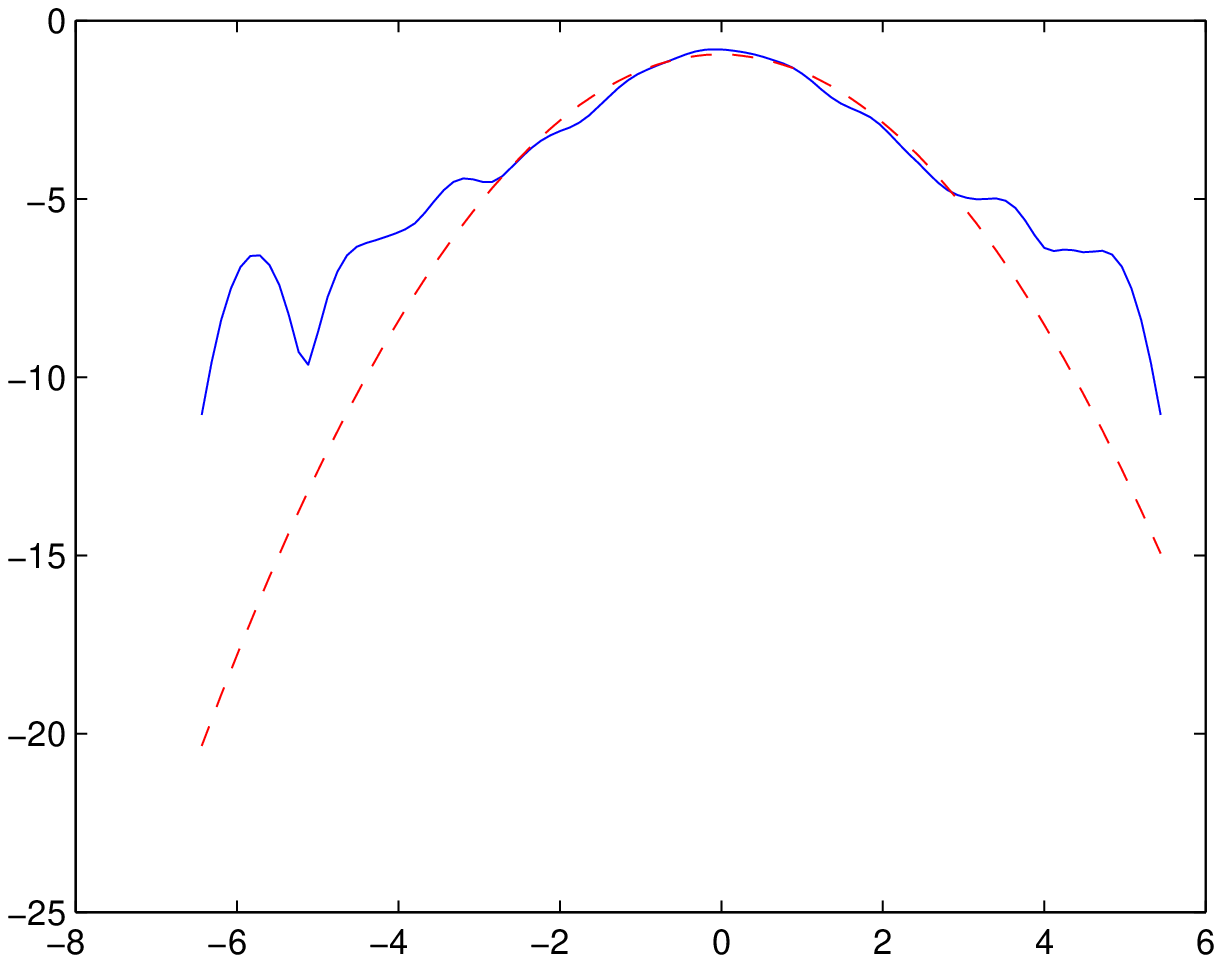} & \includegraphics[width=7.3cm,height=6.4cm]{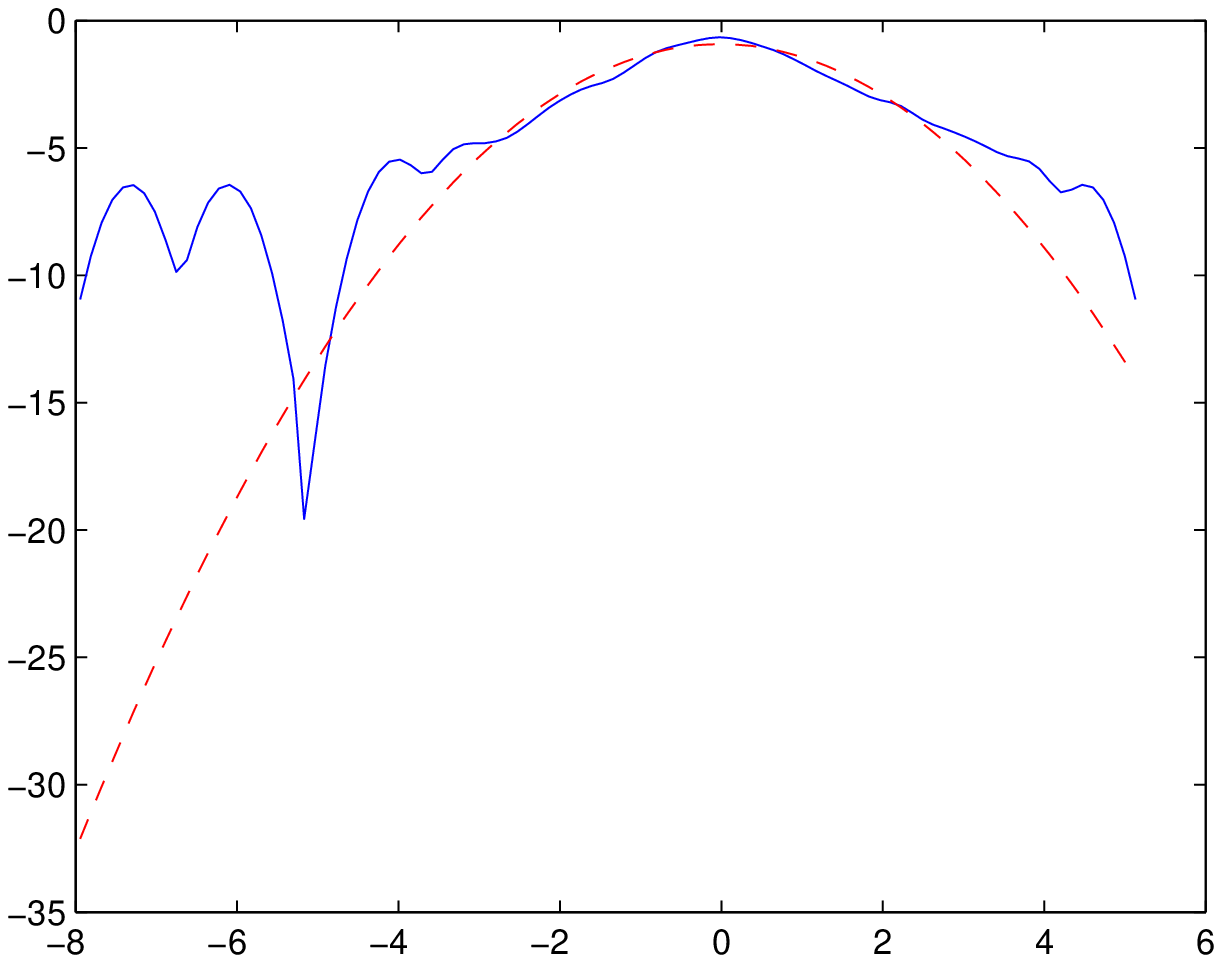}
\end{array}
$$
\caption{Theoretical log densities (dashed line) and kernel
estimates of the empirical ones (solid line) of log returns for
IBM (\emph{left}) and MSFT (\emph{right}).\label{logdensityr}}
\end{figure}
\section{Conclusion and further developments}
We introduce a new variant of the Barndorff-Nielsen and Shephard
stochastic volatility model where the non Gaussian
Ornstein-Uhlenbeck process describes some measure of trading
intensity like trading volume or number of trades instead of
unobservable instantaneous variance.

This allows us to implement a martingale estimating function
approach and obtain an explicit consistent and asymptotically
normal estimator. We first perform the numerical finite sample
experiment to assess the quality of our procedure and then apply
the obtained results to real stock data.

According to the residual analysis and to the return distribution,
the model fit is in many aspects quite satisfactory except for the
autocorrelation function of the trading volume. The graph
indicates that superposition of OU-processes could be used for a
more accurate description of the autocorrelation function, see
also \cite{Lin2007,GS2006}, but it is not clear how to extend the
martingale estimating function approach in this direction. This is
left open for future research.

In this paper the empirical analysis uses trading volume. It would
be interesting to compare the results to a similar analysis using
number of trades as suggested by \cite{Lin2007}.

The present analysis was performed for daily data, but the
approach applies for any sampling frequency since it is based on
the continuous time specification of the Barndorff-Nielsen and
Shephard model. In particular, the approach could be applied
directly to high frequency data.

Further and alternative developments like optimal quadratic
estimating functions, use of trigonometric moments and comparison
to the generalized method of moments suggested in \cite{HP2007}
apply also to the present framework.

\appendix
\section{Some numerical and analytical aspects of the density function of log returns}\label{App}
In this section we compute and analyze in detail the distribution
and the density function of log returns $\dis (X_i)_{i\geq 1}$. By
stationarity, it is sufficient to show the results for $X_1$. The
main tool for the computation is the well-known key formula, see
for example \cite{NV,ER}.
\subsection{Cumulant of $(Z_1,U_1)$}\label{KZ1U1}
It is convenient to introduce the bivariate cumulant function
$$k_{Z_1,U_1}(h_1,h_2)=K[h_1,h_2\ddagger Z_1,U_1].$$
Using relation (\ref{defuv}) and the key formula it easily follows
that
$$k_{Z_1,U_1}(h_1,h_1)=\log
E\big[\exp\{h_1Z_1+h_2U_1\}\big]=\lambda\int_{0}^{\Delta}k_{Z_1}(h_1+h_2\exp\{-\lambda(\Delta-s)\})ds,$$
where $\dis k_{Z_1}$ is the cumulant function of $Z_1.$ Moreover,
$\dis k_{Z_1}$ is explicit and related with any self-decomposable
law through the well-know formula given in \cite{BNS}.
\begin{remark}In the two concrete specifications given in sections~\ref{Sec-GaOU} and~\ref{Sec-IGOU} , namely the
$\Gamma$-OU and the IG-OU, the cumulant function of $Z_1$ is
$$k_{Z_1}(h)=\frac{\nu h}{\alpha-h}\qquad\mathrm{and}\qquad
k_{Z_1}(h)=h\delta(\gamma^2-2h)^{-1/2}$$ respectively.
\end{remark}
\subsection{Cumulant of $(Y_1,Z_1)$}\label{KY1Z1}
It is convenient to introduce the bivariate cumulant function
$$k_{Y_1,Z_1}(h_1,h_2)=K[h_1,h_2\ddagger Y_1,Z_1].$$
Using relations (\ref{deft}), (\ref{IV}), the key formula and the
independency of $Z_1$ and $\tau_0,$ it follows that
\begin{eqnarray}k_{Y_1,Z_1}(h_1,h_2)&=&\log
E\big[\exp\{h_1Y_1+h_2Z_1\}\big]\nonumber\\
&=&
k_{\tau_0}(h_1\cdot\epsilon_\lambda(\Delta))+\lambda\int_0^{\Delta}k_{Z_1}
(h_1\cdot\epsilon_\lambda(\Delta-s)+h_2)ds,
\end{eqnarray}
where $\dis k_{\tau_0}$ is the cumulant function of the trading
volume/number of trades process.
\subsection{Cumulant of $X_1$}
Finally, we are able to calculate the cumulant function of log
returns according to the obtained expressions of bivariate
cumulants in sections~\ref{KZ1U1} and~\ref{KY1Z1}. Furthermore,
using relation (\ref{defxd}) and the key formula it follows that
\begin{eqnarray}k_{X_1}(h)&=&\log E\big[\exp\{hX_1\}\big]\nonumber\\
&=&\log E\big\{ E\big[\exp\{h\mu\Delta+h\beta
Y_1+h\sigma\sqrt{Y_1}W_1+h\rho Z_1\}|Z_1,Y_1\big]\big\}\nonumber\\
&=& \log E\big[\exp\{h\mu\Delta+\beta hY_1+\rho h
Z_1+\frac{\sigma^2 h^2Y_1}{2}\}\big]\\
&=& h\mu\Delta+k_{Z_1,Y_1}(\beta h+\frac{\sigma^2 h^2}{2},\rho
h)\nonumber\\
&=& h\mu\Delta+k_{\tau_0}\big( (\beta h+\frac{\sigma^2
h^2}{2})\cdot\epsilon_\lambda(\Delta)\big)+\lambda\int_0^{\Delta}k_{Z_1}\big((\beta
h+\frac{\sigma^2 h^2}{2})\cdot\epsilon_\lambda(\Delta-s)+\rho
h\big)ds.\nonumber
\end{eqnarray}
Since in the $\Gamma$-OU case we have that
$$k_{\tau_0}(h)=\nu\log\frac{\alpha}{\alpha-h}\qquad\mathrm{and}\qquad k_{Z_1}(h)=\frac{\nu
h}{\alpha-h},$$ integrating out the cumulant function of $Z_1,$ it
follows that
\begin{eqnarray}\label{KX1} k_{X_1}(h)&=&
h\mu\Delta+\nu\log\frac{\alpha}{\alpha-\epsilon_\lambda(\Delta)\big(\beta
h+\frac{\sigma^2
h^2}{2}\big)}+\lambda\int_0^{\Delta}k_{Z_1}\big((\beta
h+\frac{\sigma^2 h^2}{2})\cdot\epsilon_\lambda(\Delta-s)+\rho
h\big)ds\nonumber\\
&=& h\mu\Delta
+\nu\log\frac{2\alpha}{2\alpha-\epsilon_\lambda(\Delta)(2\beta+\sigma^2h)}\\
&
&+\frac{\lambda\nu\bigg\{h\Delta(2\beta+2\lambda\rho+\sigma^2h)+2\alpha\log\bigg[\dis\frac{\alpha\lambda-\frac{1}{2}\lambda
h\big(2\beta\epsilon_\lambda(\Delta)+2\rho+\epsilon_\lambda(\Delta)\sigma^2
h\big)}{\lambda(\alpha-\rho
h)}\bigg]\bigg\}}{2\alpha\lambda-h(2\beta+2\lambda\rho+\sigma^2
h)}\nonumber.
\end{eqnarray}
Furthermore, the density function of log returns will be
calculated by Laplace inversion. Denoting the density function of
log returns by $f_X(x),$ we have
\begin{equation}\label{DX1}f_{X_1}(x)=\frac{1}{\pi}\int_0^{\infty}Re(\exp\{k_{X_1}(iy)-ixy\})dy,
\end{equation}where $Re(\cdot)$ denotes the real part of a complex
number. The numerical integration is performed in MATLAB using the
function \texttt{quadgk}.
\bibliography{siest1}
\end{document}